\newcommand*{\ATLASLATEXPATH}{}
\documentclass[PAPER, texlive=2016, UKenglish, cernpreprint]{\ATLASLATEXPATH atlasdoc}

\usepackage{\ATLASLATEXPATH atlaspackage}
\usepackage{float}
\floatstyle{plaintop}
\restylefloat{table}
 
\usepackage{\ATLASLATEXPATH atlasbiblatex}
 
\usepackage{\ATLASLATEXPATH atlasphysics}
\usepackage{\ATLASLATEXPATH atlasunit}
\usepackage{\ATLASLATEXPATH atlasheavyion}
\usepackage{\ATLASLATEXPATH atlasjetetmiss}

\usepackage{multirow}
 
\graphicspath{{logos/}{figures/}}
 
\usepackage{ANA-HION-2018-20-PAPER-defs}

\AtlasTitle{Transverse momentum and process dependent azimuthal anisotropies in $\sqrt{s_{\mathrm{NN}}}=8.16$~TeV $p$+Pb collisions with the ATLAS detector}
 
\AtlasAbstract{
The azimuthal anisotropy of charged particles produced in
$\sqrt{s_{\mathrm{NN}}}=8.16$~TeV $p$+Pb collisions is measured with
the ATLAS detector at the LHC. The data correspond to an integrated
luminosity of $165$~$\mathrm{nb}^{-1}$ that was collected in
2016. Azimuthal anisotropy coefficients, elliptic $v_2$ and
triangular $v_3$, extracted using two-particle correlations with a
non-flow template fit procedure, are presented as a function of
particle transverse momentum ($p_\mathrm{T}$) between $0.5$ and
$50$~GeV. The $v_2$ results are also reported as a function of
centrality in three different particle $p_\mathrm{T}$ intervals. The
results are reported from minimum-bias events and jet-triggered
events, where two jet $p_\mathrm{T}$ thresholds are used. The
anisotropies for particles with $p_\mathrm{T}$ less than about
$2$~GeV are consistent with hydrodynamic flow expectations, while
the significant non-zero anisotropies for $p_\mathrm{T}$ in the
range $9$--$50$~GeV are not explained within current theoretical
frameworks. In the $p_\mathrm{T}$ range $2$--$9$~GeV, the
anisotropies are larger in minimum-bias than in jet-triggered
events. Possible origins of these effects, such as the changing
admixture of particles from hard scattering and the underlying
event, are discussed.  }

\AtlasRefCode{HION-2018-20}
 
\PreprintIdNumber{CERN-EP-2019-217}
 
\AtlasDate{\today}

\AtlasJournal{EPJ C}
\AtlasJournalRef{Eur. Phys. J. C 80 (2020) 73}
\AtlasDOI{10.1140/epjc/s10052-020-7624-4}

\hypersetup{pdftitle={ATLAS document},pdfauthor={The ATLAS Collaboration}}
 
\begin{document}
 
\maketitle

\section{Introduction}
 
The collisions of heavy nuclei at relativistic speeds generate hot and dense droplets
of matter composed of deconfined quarks and gluons known as the quark--gluon plasma
(QGP)~\cite{Heinz:2013th,Muller:2012zq}. Studies of the QGP at the Relativistic Heavy Ion
Collider (RHIC) and the Large Hadron Collider (LHC) have yielded a wealth of surprising
results that reveal a complex set of QGP-related phenomena. Bulk hadron production, occurring
mainly at low transverse momentum ($\pt \lesssim 3$~\GeV), exhibits
significant azimuthal anisotropies that are well described in terms of nearly inviscid
hydrodynamic flow of the QGP~\cite{Heinz:2013th}. The final hadron momentum
anisotropies arise from inhomogeneities in the initial spatial distribution of the QGP
translated to momentum space via strong differential pressure gradients. These
anisotropies are characterised in terms of a Fourier decomposition:
 
\begin{equation*}
Y(\phi) = G\Big[1 + 2\sum^{\infty}_{n=1} v_{n} \cos(n(\phi-\Psi_{n}))\Big],
\end{equation*}
where $v_{n}$ are the anisotropy coefficients, $\Psi_{n}$ is the
$n^\mathrm{th}$-order orientation of the anisotropy, and the normalization, $G$,
is set by the integral of the distribution. In particular, \vtwo{} and \vthree{}
are referred to as the elliptic and triangular coefficients, respectively.
 
In addition, the production of high transverse momentum hadrons ($\pt \gtrsim 10$~\GeV) is
highly suppressed relative to the yields expected from nuclear thickness scaling of
proton--proton collision yields~\cite{Connors:2017ptx}. This suppression is understood
to result from high momentum transfer parton--parton interactions followed by the
outgoing partons losing energy via radiative and collisional processes in the QGP --
processes referred to as jet
quenching~\cite{Qin:2015srf,Mehtar-Tani:2013pia,Blaizot:2015lma}. These high-\pT
hadrons and associated jets are also observed to have a non-zero azimuthal
anisotropy~\cite{CMS-HIN-15-014,HION-2012-12,Acharya:2018lmh}, despite being well
outside the nominal domain where the anisotropies are interpreted in terms of
hydrodynamic flow. Instead, these anisotropies are understood to also arise from
inhomogeneities in the initial spatial distribution of the QGP, but in this case, where
the jet quenching effect is stronger for partons traversing longer paths through the
QGP and weaker for partons traversing shorter paths~\cite{Gyulassy:2000gk}. In this
way, low- and high-\pT hadrons have a common orientation of their azimuthal anisotropy
in a given event, because both are correlated with the orientation of the initial
geometry of the colliding nucleons. It is notable that, for more than a decade, an outstanding
and challenging theoretical puzzle has been how to quantitatively describe both high-\pT hadron suppression and azimuthal anisotropy in \PbPb collisions~\cite{Shuryak:2001me}.
There are a number of proposed explanations for resolving this puzzle in heavy-ion
collisions -- see Refs.~\cite{Molnar:2013eqa, Noronha-Hostler:2016eow, Zhang:2013oca, Liao:2008dk, Renk:2014nwa, Kopeliovich:2012sc} for examples.
 
Measurements in smaller collision systems, \pp and \pPb{} collisions at the
LHC~\cite{HION-2012-13, HION-2016-01, HION-2015-09, CMS-QCD-10-002, CMS-HIN-14-006, Abelev:2013bla}
and \pAu, \dAu, and \heAu at RHIC~\cite{PHENIX:2018lia}, indicate significant azimuthal
anisotropies for low-\pT hadrons with patterns quite similar to those observed in the
larger heavy-ion collision systems. For a recent review see
Ref.~\cite{Nagle:2018nvi}. These observations have raised the question of whether
smaller and shorter-lived droplets of QGP are formed in these smaller collision
systems. Indeed, models employing nearly inviscid hydrodynamics for the QGP provide a
quantitative description of this large body of data at low \pT~\cite{Romatschke:2017ejr}.
 
In contrast, measurements aimed at observing signatures of jet quenching in small
collision systems have found no such effect. Measurements of hadron and jet \pt spectra
at high \pT indicate production yields consistent with those in \pp collisions scaled
up by the expected nuclear thickness in \pPb~\cite{HION-2013-08,CMS-HIN-14-001,ALICE:2012mj} and
$d$+Au collisions~\cite{Adare:2015gla}, and that the \pt-balance between dijets or
hadron--jet pairs is unmodified in \pPb collisions within
uncertainties~\cite{CMS-HIN-13-001,Acharya:2017okq}. The ATLAS experiment has also
published results for the hadron azimuthal anisotropy up to $\pt \approx 12$~\GeV{}
that hint at a non-zero anisotropy extending into the region beyond the
usual hydrodynamic interpretation and into the regime of jet
quenching~\cite{HION-2013-04}. However, it is unlikely that there can be differential
jet quenching as a function of orientation relative to the QGP geometry if there
is no jet quenching in \pPb collisions as observed in the spectra. Thus, there are two related
outstanding puzzles, one being the lack of jet quenching observed in the spectra, if indeed
small droplets of QGP are formed, and the other being what mechanism can lead to high-\pT
hadron anisotropies other than differential jet quenching.
 
This paper presents a measurement of the azimuthal anisotropy of unidentified hadrons
as a function of \pt and centrality in $\sqn = 8.16$~\TeV{} \pPb collisions with the
ATLAS detector. The measurement is made using two-particle correlations, measured separately
for minimum-bias triggered (MBT) events and events requiring a jet with \pt greater than either
75~\GeV{} or 100~\GeV{}. There are contributions to the azimuthal correlations from
particle decays, jets, dijets, and global momentum conservation, which have traditionally
been referred to as `non-flow'~\cite{Lim:2019cys}. Using this nomenclature, a standard
template fitting procedure is applied to subtract non-flow contributions~\cite{HION-2016-01,HION-2015-09}.
To decrease the residual influence of the non-flow correlation in the jet events, a
novel procedure is used to restrict the acceptance of particles according to the location
of jets in the event. Assuming that the two-particle anisotropy coefficients are the
products of the corresponding single-particle coefficients (factorisation), the
elliptic and triangular anisotropy coefficients, \vtwo{} and \vthree{}, are reported as
a function of \pt. Additionally, \vtwo{} results are presented as a function of centrality in
three different \pt ranges. Finally, the fractional contribution to the correlation
functions from jet particles is determined as a function of \pt.

\section{ATLAS detector}
 
The ATLAS experiment~\cite{PERF-2007-01} at the LHC is a multipurpose particle
detector with a forward--backward symmetric cylindrical geometry and
nearly $4\pi$ coverage.\footnote{ATLAS uses a right-handed coordinate
system with its origin at the nominal interaction point (IP) in the
centre of the detector and the $z$-axis along the beam pipe. The
$x$-axis points from the IP to the centre of the LHC ring, and the
$y$-axis points upward.  Cylindrical coordinates $(r, \phi)$ are
used in the transverse plane, $\phi$ being the azimuthal angle
around the $z$-axis. The pseudorapidity is defined in terms of the
polar angle $\theta$ as $\eta = -\ln \tan(\theta/2)$. Transverse
momentum and transverse energy are defined as $\pT = p \sin\theta$
and $\ET = E \sin\theta$, respectively. Angular distance $\Delta{R}$ is defined as
$\sqrt{ (\Delta\eta)^2 + (\Delta\phi)^2 }$.} This analysis relies on
the inner detector, the calorimeter, and the data acquisition and
trigger system.
 
The inner detector (ID) comprises three major subsystems: the pixel
detector and the silicon microstrip tracker, which extend up to
$\left|\eta\right| = 2.5$, and the transition radiation tracker, which
extends to $\left|\eta\right| = 2.0$. The inner detector covers the
full azimuth and is immersed in a 2~T axial magnetic field. The pixel
detector consists of four cylindrical layers in the barrel region and
three discs in each endcap region. A new innermost layer, the insertable
B-layer~\cite{ATLAS-TDR-19,PIX-2018-001}, has been operating as a part of the pixel detector
since 2015. The silicon microstrip tracker comprises four cylindrical layers (nine
discs) of silicon strip detectors in the barrel (endcap) region. The minimum-bias
trigger scintillator detects charged particles over
$2.07<|\eta|<3.86$ using two hodoscopes of 12 counters positioned at $|z|=3.6$~m.
 
The calorimeter is a large-acceptance, longitudinally segmented
sampling detector covering $\left|\eta\right| < 4.9$ with
electromagnetic (EM) and hadronic sections. The EM calorimeter is a
lead/liquid-argon sampling calorimeter with an accordion-shaped
geometry. It is divided into a barrel region, covering
$\left|\eta\right| < 1.475$, and two endcap regions, covering $1.375 < \left|\eta\right| < 3.2$. The hadronic calorimeter surrounds the EM
calorimeter. It consists of a steel/scintillator-tile sampling
calorimeter covering $\left|\eta\right| < 1.7$ and a liquid-argon
calorimeter with copper absorber covering $1.5 < \left|\eta\right| < 3.2$. The forward calorimeter (FCal) is a liquid-argon sampling calorimeter
located on either side of the interaction point. It covers $3.1 < \left|\eta\right| < 4.9$ and each half is composed of one EM and two
hadronic sections, with copper and tungsten serving as the absorber
material, respectively. The FCal is used to characterise the centrality
of $p$+Pb collisions as described below.
 
In this analysis, a two-level trigger system was used to select events, with a first-level (L1) trigger implemented in hardware followed by a software-based high-level trigger (HLT) which reconstructs the event in a manner similar to the final offline reconstruction. Events used for the measurements presented in this paper were selected using several triggers. MBT events were selected by a trigger that requires a signal in at least one minimum-bias trigger scintillator counter at L1~\cite{TRIG-2016-01} followed by the requirement of at least one reconstructed track at the HLT stage. Events with a high-$p_\mathrm{T}$ jet were acquired using a high-level jet trigger covering the central region ($\left|\eta\right| < 3.2$). These events were first required to have energy deposits at L1 that are compatible with the presence of a jet and then to pass various thresholds for the jet transverse energy at the HLT stage.

\section{Data and event selection}
 
During $p$+Pb data-taking in 2016, the LHC was configured with a beam composed of protons with an energy of 6.5~\TeV\ and a beam of lead ions with an energy per nucleon of 2.51~\TeV. This resulted in a collision system with proton--nucleon centre-of-mass energy $\sqn = 8.16~\TeV$ and a rapidity shift of the centre of mass by +0.465 units in the proton-going direction relative to the laboratory frame. The data were taken over two running periods with different configurations of the LHC beam directions. In the first period of data-taking, comprising a total integrated luminosity of 57~$\mathrm{nb}^{-1}$, the lead ions circulated clockwise in beam 1, while the protons circulated counterclockwise in beam 2. For the second period of data-taking, which comprised 108~$\mathrm{nb}^{-1}$, the beam species were interchanged. The analysed data were provided by the minimum-bias trigger described above, which was prescaled and sampled 0.079~$\mathrm{nb}^{-1}$ of luminosity. In addition, data were selected by the high-level jet triggers with transverse energy thresholds of 75~\GeV\ and 100~\GeV, which sampled 26~$\mathrm{nb}^{-1}$ and the full 165~$\mathrm{nb}^{-1}$ of \pPb{} luminosity, respectively.
 
Events selected by the triggers described above were reconstructed offline following procedures that were optimised for the Run-2 detector configuration~\cite{ATL-PHYS-PUB-2015-018}.  
Events are required to have at least one reconstructed vertex. To reduce the contribution from events with multiple in-time \pPb interactions, events with more than one vertex are used only if the additional vertices have fewer than seven associated reconstructed tracks with $\pt > 0.4$~\GeV. That is, events are only allowed to have one vertex with seven or more associated tracks. Two classes of jet events were defined by requiring an offline jet with $\pt > 75$~\GeV\ or $\pt>100$~\GeV\ respectively, and were drawn from the jet-triggered event samples with the analogous online thresholds. The trigger efficiency, given this offline selection, was greater than 97\% for both jet samples.
 
Events were further characterised by the sum of the transverse energy in the FCal module
in the direction of the Pb beam, \sumETPb. The event centrality is defined as the \sumETPb
percentile of the events in minimum-bias collisions, after accounting for the
inefficiency introduced by the trigger and event selection criteria, and was determined
in a way similar to previous analyses of Run-1 $p$+Pb data at $\sqn =
5.02~\TeV$~\cite{HION-2013-08,HION-2012-15}. Events within the 0--90\% centrality range
were used in this analysis, with low (high) values corresponding to high-\sumETPb
(low-\sumETPb) events with large (small) overall particle multiplicity. Since the
acceptance of the FCal is separate from that of the ID, this centrality definition has
the benefit of reducing event-selection-induced biases in the measured
quantities~\cite{Adam:2014qja}.
 
\section{Track and jet reconstruction}
 
The reconstruction, selection, and calibration of charged-particle tracks and calorimetric jets, and their performance as determined using Monte Carlo (MC) simulations, are described below.
 
Charged-particle tracks and collision vertices are reconstructed in the ID using the algorithms described in Ref.~\cite{ATL-PHYS-PUB-2015-018}.
Inner detector tracks with $\pt>0.4$~\GeV\ and $|\eta|<2.5$ were required to satisfy a set of quality criteria similar to those described in Ref.~\cite{HION-2016-01}.
The total number of reconstructed ID tracks satisfying these selection criteria in a given event is called the multiplicity or \nch.
The reconstruction and selection efficiency for \textit{primary}~\cite{PERF-2015-07} charged hadrons to meet these criteria was determined using a sample of 3~million minimum-bias $p$+Pb events simulated by the \HIJING generator~\cite{Deng:2010mv}. Events were generated with both beam configurations. The ATLAS detector response to the generated events was determined through a full \GEANT{4} simulation~\cite{Agostinelli:2002hh,SOFT-2010-01}, and the simulated events were reconstructed in the same way as the data. Over the measured kinematic range, the efficiency varies from approximately $50\%$ for the lowest-\pt\ hadrons at large pseudorapidity, to greater than $90\%$ for hadrons with $\pt > 3$~\GeV\ at mid-rapidity.
 
Jets are reconstructed using energy deposits in the calorimeter system, $\left|\eta\right| < 4.9$, in a range partially overlapping with both the ID and the FCal used to determine centrality. The reconstruction closely follows the procedure used in other measurements for Pb+Pb and $pp$ collisions~\cite{HION-2011-02,HION-2017-04}. Jets are measured by applying the anti-$k_t$ algorithm~\cite{Cacciari:2008gp,Cacciari:2011ma} with radius parameter $R=0.4$ to energy deposits in the calorimeter. No jets with $\pt<15$~\GeV\ are considered. An iterative procedure is used to obtain an event-by-event estimate of the $\eta$-dependent underlying-event energy density, while excluding jets from that estimate. The jet kinematics are corrected for this background and for the detector response using an $\eta$- and $\pt$-dependent calibration derived from fully simulated and reconstructed \PYTHIAV{8}~\cite{Sjostrand:2007gs} hard-scattering events configured with the NNPDF23LO parton distribution function set~\cite{Ball:2012cx} and the A14 set of tuned parameters~\cite{ATL-PHYS-PUB-2014-021} to model non-perturbative effects. An additional, small correction, based on {\em in situ} studies of jets recoiling against photons, $Z$ bosons, and jets in other regions of the calorimeter, is applied~\cite{PERF-2016-04,HION-2018-06}.  
Simulation studies show that for jets with $\pt > 75$~\GeV, the average reconstructed jet \pt is within 1\% of the generator level jet \pt and has a relative \pt resolution below $10$\% after the calibration procedure.
 
\section{Analysis procedure}
\label{sec:analysis}
 
This analysis is based on previous ATLAS two-particle correlation
studies~\cite{HION-2016-01,HION-2015-09}. To construct the two-particle correlation
functions, the selected inner-detector tracks with $\pt > 0.4$~\GeV{} are divided into
two sets, referred to as A- and B-particles in this paper,
although they are sometimes referred to as trigger and associated particles in the literature.
To reduce the contribution of non-flow correlations from decays and jets, two restrictions are
placed on A--B particle pairs drawn from the two sets. First, as was done in previous analyses~\cite{HION-2016-01,HION-2015-09}, the particles are required to be separated in
pseudorapidity with $|\deta^\mathrm{AB}| = |\eta^\mathrm{B}-\eta^\mathrm{A}| > 2$. This
requirement removes the short-range decay and jet fragmentation structure, while emphasising
global, early-time correlations. Due to the enhanced contribution from jet correlations
in the jet-triggered events, an additional constraint was developed for this analysis.
Namely, B-particles are required to be separated in pseudorapidity by one unit from all
reconstructed jets with $\pt^{\mathrm{jet}}>15$~\GeV{}, i.e.\
$|\deta^{\mathrm{jB}}| = |\eta^\mathrm{B}-\eta^{\mathrm{jet}}| > 1$. This latter requirement is
only applied to the jet-triggered events, and in this way, the jets act as a source
of high-\pT A-particles but contribute few B-particles.
 
The correlation functions, $S(\dphi)$, are defined as the yields of particle pairs passing the above event and pair selection,
binned in $\dphi=\phi^\mathrm{A}-\phi^\mathrm{B}$, and normalised by the total number of A-particles.
Corrections for the imperfect trigger and tracking efficiencies are applied as weights to
the entries of the correlation functions. A mixed-event correction,
$M(\dphi)$, is generated by correlating A-particles from one event with B-particles
from a different event with a vertex $z$-position differing by less than 10~\mm{} and a
number of reconstructed charged particles (\nch{}) differing by less than 10 for $\nch<100$
and less than 20 for $\nch > 100$. Thus, the mixed events contain only trivial detector
acceptance effects and no physical correlations. To reduce the statistical uncertainty
introduced by the correction, each event is mixed with five others meeting the above vertex
$z$ and \nch{} conditions. The corrected correlation is, then, $Y(\dphi)=S(\dphi)/M(\dphi)$,
where $M(\dphi)$ is normalised such that the ratio preserves the overall integral of
$S(\dphi)$. Jet events are mixed with other jet events, and the $|\deta^{\mathrm{jB}}|$ condition
is applied with respect to the jets in the A-particle event only. Thus, the B-particle acceptance
is consistent between the same- and and mixed-event correlations.
 
\begin{figure}[!htbp]
\begin{center}
\includegraphics[width=0.44\textwidth]{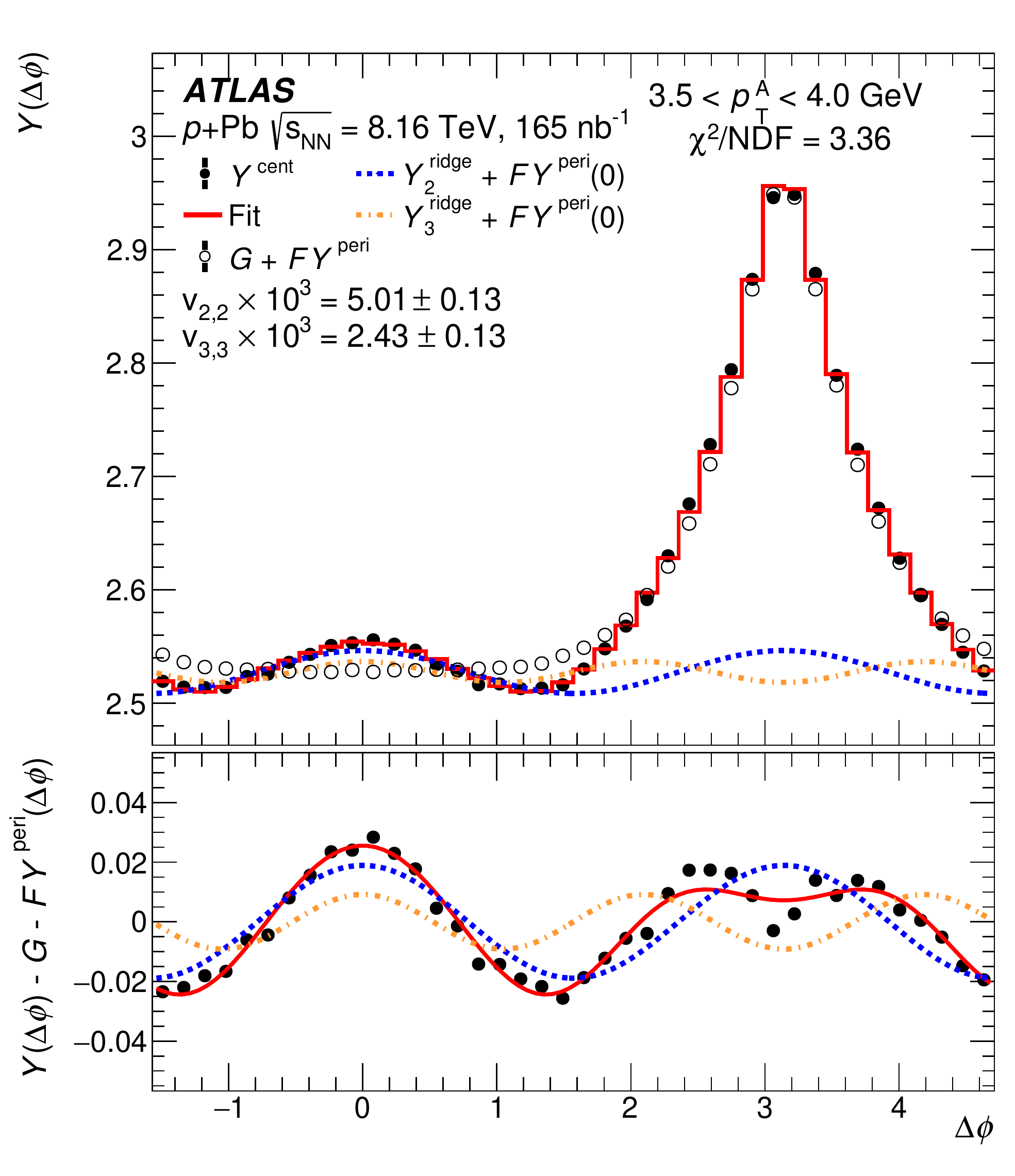}
\includegraphics[width=0.44\textwidth]{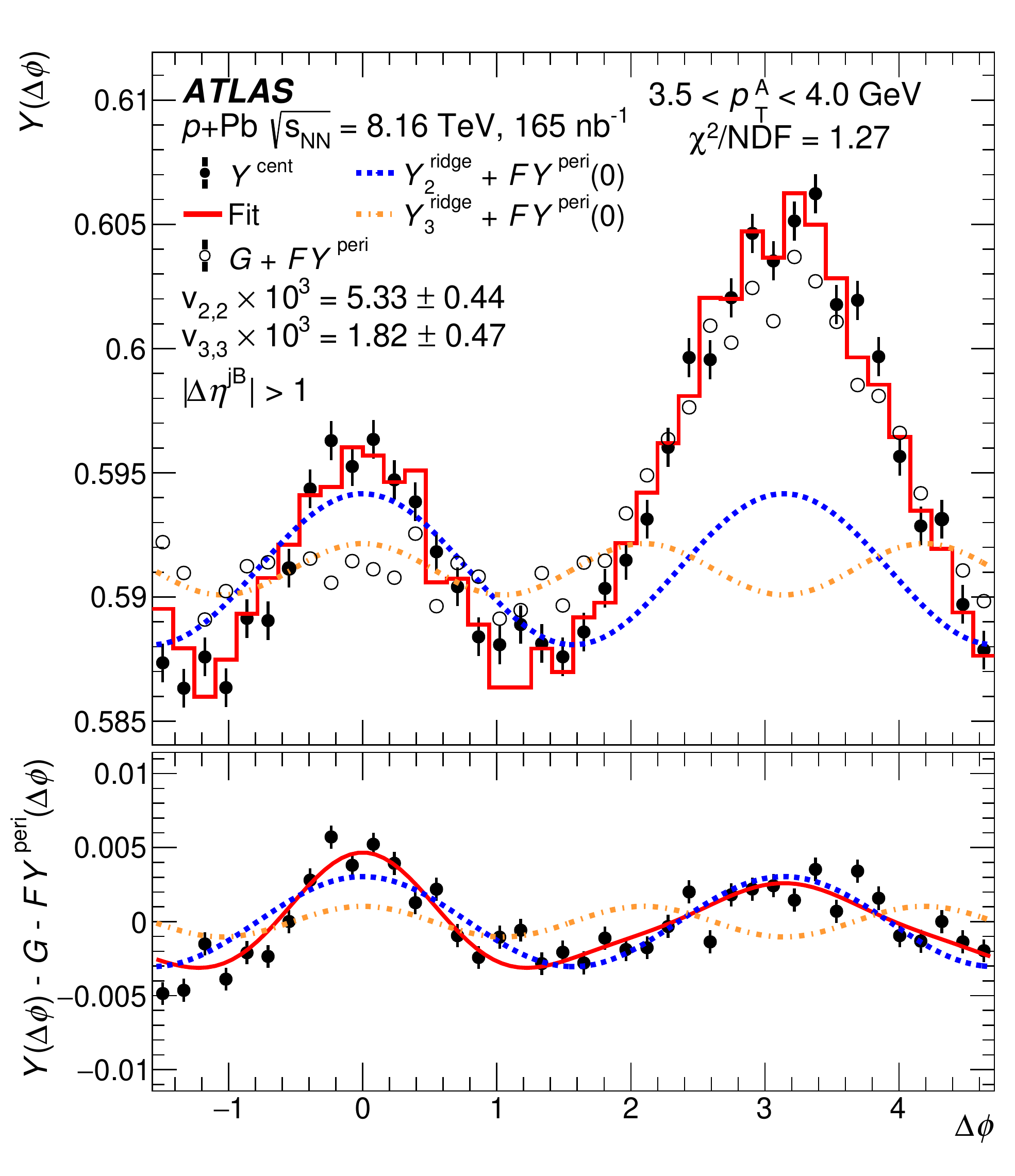}
\caption{Template fitting output for events with jet $\pt>100$~\GeV{}. Both require $3.5<\pt^\mathrm{A}<4.0~\GeV$ and are made with 60--90\% peripheral
selection and 0--5\% central selection. The left plot is made with no selection on
the B-particles and the right plot is made requiring the B-particles to have
$|\deta^{\mathrm{jB}}|>1$ relative to all jets with $\pt^\mathrm{jet} > 15~\GeV$ in the
event. In the upper panels, the open circles show the scaled and shifted peripheral
template with uncertainties omitted, the closed circles show the central
data, and the red line shows the fit (template and harmonic functions). The
blue dashed line shows the second-order harmonic component, $Y_2^\mathrm{ridge}$, and the
orange dotted line shows the third-order harmonic component, $Y_3^\mathrm{ridge}$ (the $n=2$ and
$n=3$ contributions to the sum in \Eqn{\ref{eqn:temp}}, respectively). The lower panels show the
difference between the central data and the peripheral template along with
the second and third harmonic functions. The resulting $v_{2,2}$, $v_{3,3}$, and global fit $\chi^2$/NDF
values are reported in the legends. In these fits, NDF$=35$.
}
\label{fig:fits}
\end{center}
\end{figure}

\begin{figure}[!htbp]
\begin{center}
\includegraphics[width=0.44\textwidth]{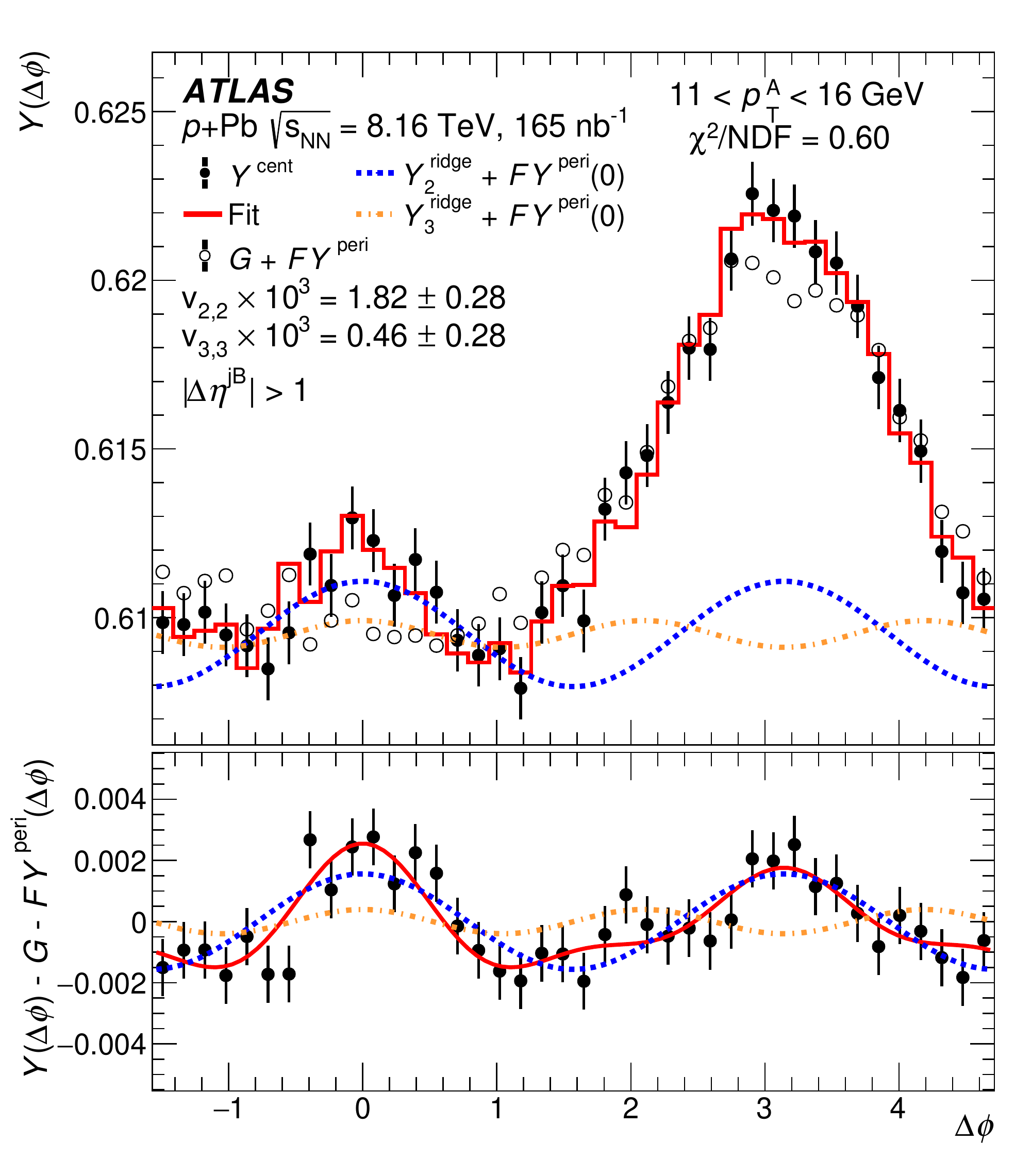}
\includegraphics[width=0.44\textwidth]{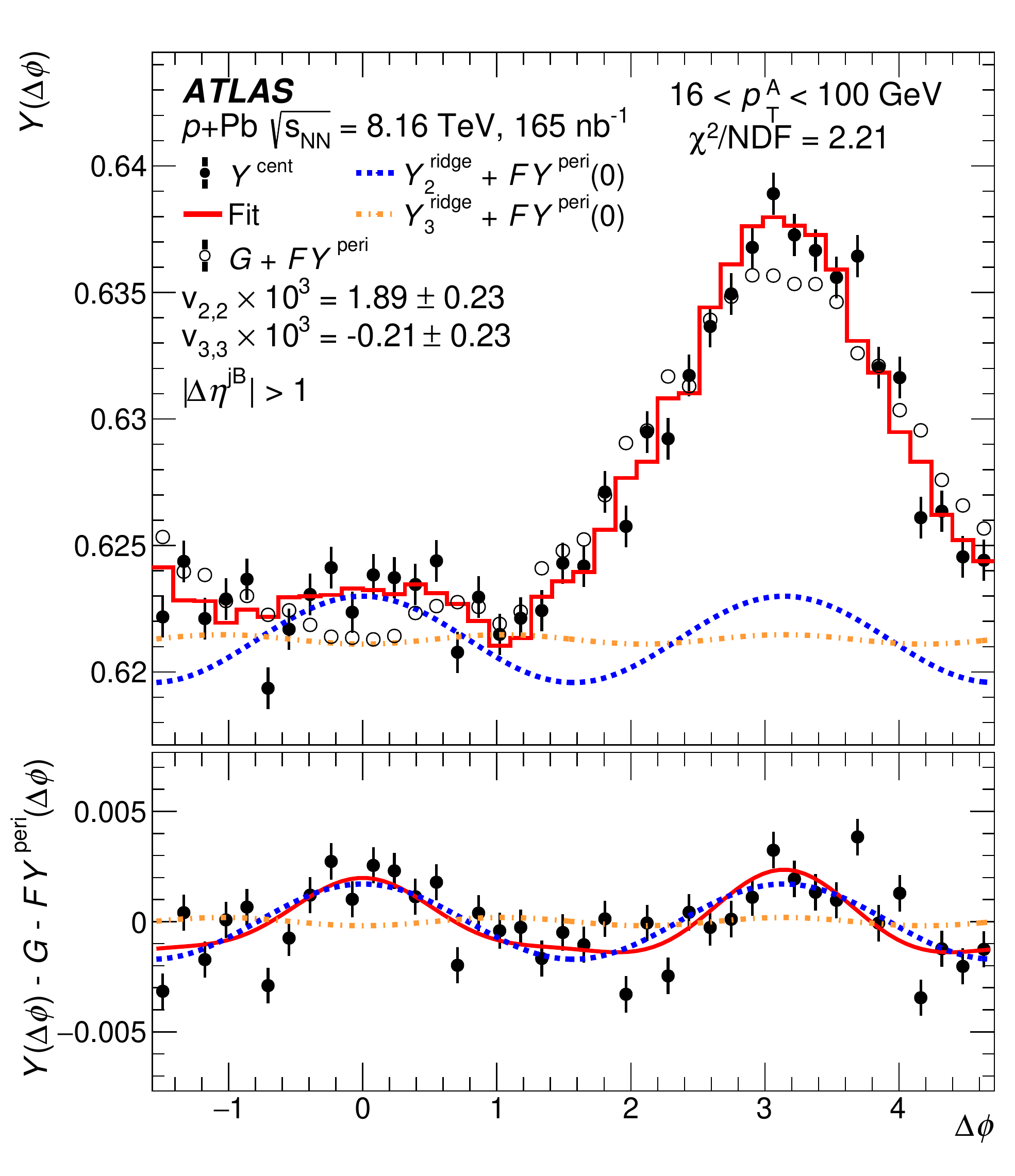}
\caption{Template fitting output for events with jet $\pt>100$~\GeV{} with
60--90\% peripheral selection and 0--5\% central selection. The left plot is made with
$11<\pt^\mathrm{A}<16~\GeV$ and the right plot with $16<\pt^\mathrm{A}<100~\GeV$. In the upper
panels, the open circles show the scaled and shifted peripheral template with
uncertainties omitted, the closed circles show the central data, and the red line
shows the fit (template and harmonic functions). The blue dashed line shows the
second-order harmonic component, $Y_2^\mathrm{ridge}$, and the orange dotted line
shows the third-order harmonic component, $Y_3^\mathrm{ridge}$ (the $n=2$ and $n=3$
contributions to the sum in \Eqn{\ref{eqn:temp}}, respectively). The lower panels show
the difference between the central data and the peripheral template along with the second
and third harmonic functions. The resulting $v_{2,2}$, $v_{3,3}$, and global fit $\chi^2$/NDF
values are reported in the legends. In these fits, NDF$=35$.
}
\label{fig:fits2}
\end{center}
\end{figure}

To extract the anisotropy coefficients while accounting for residual non-flow, the ATLAS
template fitting procedure, as used for previous
results~\cite{HION-2016-01,HION-2015-09}, is applied to $Y(\dphi)$. In this procedure,
$Y(\dphi)$ is found for two different selections of event activity quantified by
centrality: a central selection, $Y^{\mathrm{cent}}(\dphi)$, and a peripheral
selection, $Y^{\mathrm{peri}}(\dphi)$. In this analysis, the peripheral selection corresponds to
the 60--90\% centrality interval. Assuming that the shape of the non-flow correlation is independent
of centrality, $Y^{\mathrm{cent}}(\dphi)$ is expressed as
\begin{equation}
Y^\mathrm{cent}(\Delta\phi) =  FY^\mathrm{peri}(\Delta\phi) +  G\Big[1 + 2\sum^{4}_{n=2} v_{n,n} \cos(n\dphi) \Big]\,,\\
\label{eqn:temp}
\end{equation}
where $F$ and each \vnn{} are parameters of a global $\chi^2$ fit, and $G$ is fixed by
the requirement that the integral of the fit model is that of $Y^\mathrm{cent}$. The parameter
$F$ allows for a linear scaling of the non-flow between the two centrality classes. The fit
includes the fourth harmonic, $v_{4,4}$, but it is not presented in the
results because it is statistically insignificant. The fit $\chi^2$ function incorporates the
statistical uncertainties from both $Y^{\mathrm{cent}}$ and the peripheral template, $Y^{\mathrm{peri}}$, although the examples shown in Figures~\ref{fig:fits} and~\ref{fig:fits2}
do not show the uncertainties of $Y^{\mathrm{peri}}$ for readability. The statistical
uncertainties of the extracted \vnn{} parameters are returned from the MINUIT $\chi^2$ minimiser~\cite{JAMES1975343},
accounting for correlations between parameters.
 
\Fig{\ref{fig:fits}} shows an example of two template fits using jet-triggered events with
jet $\pt > 100$~\GeV. The left plot shows the fit for correlations made
without the B-particle jet rejection condition, and the right plot shows the same
correlation, but with the condition $|\deta^{\mathrm{jB}}| > 1$ applied.
In this figure, $Y_N^\mathrm{ridge}$ represents the
$N^{\mathrm{th}}$-order harmonic component of the fit. The left plot has a dominant non-flow
contribution, and a distortion in the resulting subtracted distribution is observed near
$\Delta\phi \approx \pi$. Removing much of the jet correlation in this way reduces the overall
sensitivity to the template method assumption that the shape of the non-flow contribution is the
same for the central and peripheral selections. However, violation of this assumption will
introduce distortions that could potentially bias the harmonic coefficients. This is
explicitly tested by varying the centrality selection of the peripheral template, as
discussed further in Section~\ref{sec:sys}. Two additional examples of template fits from
the jet-triggered events with jet $\pt > 100$~\GeV\ and with the B-particle jet rejection are plotted in
\Fig{\ref{fig:fits2}}. These show the behaviour of the template fits for high A-particle \pt.
 
If the particle momentum correlations originate from a global field, as is the case
for collective expansion, the \vnn{} will factorise such that
$\vnn(\pt^\mathrm{A},\pt^\mathrm{B}) = v_n(\pt^\mathrm{A})\cdot v_n(\pt^\mathrm{B})$.
By assuming this relation and making specific \pt selections on A- and B-particles,
the single-particle $v_n(\pt^\mathrm{A})$ can be obtained from
\begin{equation*}
v_n(\pt^\mathrm{A})=\vnn(\pt^\mathrm{A},\pt^\mathrm{B})/ \sqrt{\vnn(\pt^\mathrm{B},\pt^\mathrm{B})},
\end{equation*}
where $\vnn(\pt^\mathrm{A},\pt^\mathrm{B})$ is determined with A- and B-particles
having \pt in range $\pt^\mathrm{A}$ and $\pt^\mathrm{B}$, respectively, and
$\vnn(\pt^\mathrm{B},\pt^\mathrm{B})$ is determined with A- and B-particles both
having \pt in range $\pt^\mathrm{B}$. In this analysis, this range is nominally
$\pt^\mathrm{B}>0.4$~\GeV{}, although the dependence of the extracted anisotropy on this
choice is explored in Section~\ref{sec:result}.
 
The relative yield of particle pairs entering the correlation functions is estimated
assuming a simple, two-component model of particle production. Particles are assumed to
be produced either by hard scattering (HS) processes, such as jet production, or by soft underlying
event (UE) processes. With this assumption, the correlation functions are constructed
from pairs pulled from a mixture of the two sources. Particle pairs can be formed in
the following four A--B combinations: UE--UE, UE--HS, HS--UE, and HS--HS.
The event-by-event yields of the UE and HS processes are estimated by classifying the charged
particles according to their azimuthal orientation relative to the leading jet or,
in the case of MBT events that contain no jets with $\pt > 15$~\GeV{} and $|\eta|<4.9$,
relative to the leading hadron. The following regions are defined relative to this leading
vector:
\begin{itemize}
\item towards: $(|\phi_\mathrm{B} - \phi_\mathrm{jet}| < \frac{\pi}{4}) \: \cup \: (|\phi_\mathrm{B} - \phi_\mathrm{jet}| > \frac{3\pi}{4})$
\item transverse: $\frac{\pi}{4} < |\phi_\mathrm{B} - \phi_\mathrm{jet}| < \frac{3\pi}{4}$.
\end{itemize}
Then, assuming that HS particles are completely contained in the towards
region and the UE particles are distributed uniformly in azimuth,
the following relations are inferred:
\begin{align*}
N_\mathrm{UE} & = 2 N_\mathrm{trans},\\
N_\mathrm{HS} & = N_\mathrm{toward} - N_\mathrm{trans},
\end{align*}
where $N_\mathrm{UE}$ and $N_\mathrm{HS}$ are the single-particle yields from UE
and HS processes, respectively, and $N_\mathrm{trans}$ and $N_\mathrm{toward}$ are the
particle yields in the transverse and toward regions, respectively. The quantities
$N_\mathrm{UE}$ and $N_\mathrm{HS}$ are statistically determined from the event averaged
$N_\mathrm{trans}$ and $N_\mathrm{toward}$ and are, thus, insensitive to event-by-event
fluctuations. However, it is not possible to classify individual particles. It should be noted that
the assumptions used in this derivation are likely not perfect; for example, the
UE is not uniformly distributed in $\phi$, event by event, due to the presence of
azimuthal anisotropy. The leading object may be more likely to be oriented
with the anisotropy, in which case the UE yield would be underestimated and the
HS yield overestimated. However, the analysis proceeds with the assumptions as given and
includes no additional uncertainty for this potential effect.
 
The $\deta^\mathrm{AB}$ and $\deta^\mathrm{jB}$ rejections produce a geometric correlation
between the yields of A- and B-particles and, thus, the number of pairs is not the simple
product of the two individual yields. Accounting for the dependencies, the total yield
of particle pairs can be expressed in the following way
 
\begin{align}
Y_\textrm{X--Z}=\int_{-2.5}^{2.5} \frac{\mathrm{d}N^\mathrm{A}_\mathrm{X}(\eta^\mathrm{A})}{\mathrm{d}\eta^\mathrm{A}}  \Bigg[  \int_2^5 \frac{\mathrm{d}^2N^\mathrm{B}_\mathrm{Z}(\eta^\mathrm{A},|\deta^\mathrm{AB}|)}{\mathrm{d}\eta^\mathrm{A} \mathrm{d}|\deta^\mathrm{AB}|} \mathrm{d}|\deta^\mathrm{AB}|  \Bigg] \mathrm{d}\eta^\mathrm{A} \, ,
\label{eq:pairfrac}
\end{align}
where X--Z could be any pairwise combination of UE and HS. In the case of jet events,
the $\deta^\mathrm{jB}$ condition is enforced when filling the $\eta^\mathrm{A}$ and
$\deta^\mathrm{AB}$ distributions so it's effects are taken into account.

\section{Systematic uncertainties}
\label{sec:sys}

The systematic uncertainties fall into two categories: those associated with both the MBT
and jet events and those associated with only the jet events. The uncertainties are
determined by assessing the difference between the nominal value of \vtwo{} or \vthree{} and the
value after a given variation. Unless otherwise stated, the uncertainties are defined as
asymmetric one-standard-deviation errors. The final uncertainty is the quadrature sum of
the uncertainty from each individual source. The relative downward and upward systematic
uncertainties from different sources and all sources combined are shown for \vtwo{} in
\Tab{\ref{tab:my_label}} and for \vthree{} in \Tab{\ref{tab:v3}}.
The rest of this section focuses primarily on the systematic uncertainties of \vtwo{}. While the absolute uncertainties in \vtwo{} and \vthree{} are of similar magnitude, this represents larger relative uncertainties in the \vthree{} values since they are generally smaller than the \vtwo{} values at any given \pt.

For both the MBT and jet events, the sensitivity to the trigger and tracking efficiency corrections
was assessed by removing each. This variation (`Track/trig Eff.' in Table~\ref{tab:my_label})
yields a 0--2\% (2--4\%) difference for MBT (jet) events, depending on track \pt, and is
subdominant. In the construction of the correlation functions, the uncertainty in the
mixed-event correction was again found by removing it from the analysis.
This variation results in an uncertainty that vanishes at low \pt but grows to 20\% (10\%)
at high \pT for MBT (jet) events, but remains subdominant to statistical uncertainties over
the whole \pt range. Regarding the template fitting procedure, the centrality range for the
peripheral reference selection was varied from the nominal 60--90\% to 50--70\% and 70--90\%.
This variation (`Peri.\ reference') results in an uncertainty of about 2\% at low \pt and
increasing to about 10\% or 18\% in the mid \pt range between 2 and 10~\GeV{} depending on
the event trigger. This last uncertainty is dominant in this category for most of the \pt range probed
in the measurement. At high \pt, the sensitivity of the measurements in MBT events to the mixed event correction and reference selection is significantly higher than in jet events; this is particularly noticeable for the \vthree{} values, where the relative uncertainties in the MBT events for $\pt > 10$~\GeV\ are $5$--$10$ times larger than in the jet-triggered events.
 
The following set of uncertainties is associated with jet events only. To assess the
sensitivity to the uncertainty in the jet energy scale and the impact of imperfect
trigger efficiency, the jet \pt thresholds used to select events were  varied from 75~\GeV{}
and 100~\GeV{} to 80~\GeV{} and 105~\GeV{}, respectively. This variation (`Trig jet \pt')
results in a symmetric uncertainty that is smaller than 1\% at low particle \pt and that
increases to about 10\% with increasing \pT. It is subdominant to other sources in this category.
The jets used in the B-particle jet rejection were varied to include only jets with \pt
greater than 20~\GeV{} instead of the nominal 15~\GeV{} (`Reject jet \pt'). The 2\% and 10\%
differences at low and high \pt are incorporated as a symmetric uncertainty that is subdominant
to other sources in this category. The $\deta^\mathrm{jB}$ rejection
allows jets to be composed of only a single particle that may originate in the tail of the UE
particle \pt spectrum. Thus, the jets used in this rejection were varied to require at least three
tracks in a $\Delta R = 0.4$ cone around the jet axis (`Reject jet mult.'). The uncertainty
associated with this variation is about 5\% and subdominant to the others in this category.
An additional uncertainty is used to cover the impact of a sector of the hadronic endcap
calorimeter (HEC) being disabled for the running period. The disabled sector was in the range
$1.5<\eta<3.2$ and $-\pi<\phi<-\pi/2$. This uncertainty (`Disabled HEC sector')
was assessed by requiring all B-particles to be outside the pseudorapidity region of the
disabled HEC. The difference was found to be less than 1\% at low \pt and about
20\% at high \pt, where it is the dominant positive uncertainty. Finally, an uncertainty
is assigned to account for the potential of the UE to bias the event selection. The
azimuthal modulation of the UE increases the reconstructed \pt of jets aligned with the
flow orientation, and, thus, the event-wise jet-\pt{} threshold will bias the events to have
more jets correlated with the flow plane. The impact of this effect on the measured results
was assessed in simulation by mixing jet events with a realistic UE containing azimuthal
anisotropy. The resulting uncertainty only affects \vtwo{}, is the
dominant negative uncertainty for track \pt{} greater than 3~\GeV, and is about 30\% (20\%) for
jet-triggered events with jet $\pt > 75$~\GeV\ (100~\GeV). The effect is larger for
lower-\pt jets because the UE energy contribution is independent of jet energy. For a
power-law spectrum, a given threshold change has a greater fractional effect on the yield
for smaller values of the threshold.
 
In summary, the uncertainty in \vtwo{} from the peripheral reference selection was found to be
dominant for \pt less than 10~\GeV{} for MBT events, above which, the mixed event correction
uncertainty is dominant, and between 2 and 5~\GeV{} for jet events. The
uncertainties associated with the jet selection were found to be dominant for $\pt\gtrsim10$~\GeV{}
in jet events. The total uncertainty in MBT events ranges from ($-1\%$, $+2\%$) at low
\pt to about ($-10\%$, $+25\%$) at high \pt. For jet events, the total uncertainty
ranges from about ($-5\%$, $+4\%$) at low \pt to about ($-35\%$, $+50\%$) and ($-30\%$, $+25\%$)
at high \pt for events with jet $\pt > 75$~\GeV{} and jet $\pt > 100$~\GeV{} respectively.
 
\begin{table}[]
\centering
\begin{tabular}{|l|c|c|c|c|c|c|}
\hline
\multirow{2}{*}{Source} &  \multicolumn{2}{c|}{ $\pt < 2$~\GeV} & \multicolumn{2}{c|}{$\pt=2$--$10$~\GeV} & \multicolumn{2}{c|}{$\pt > 10$~\GeV} \\ \cline{2-7}
&  MBT & Jet & MBT & Jet & MBT & Jet \\ \hline
Track/trig Eff.    & $<+1$\% & $+2$\% & $<+1$\% & $+4$\% & $-2$\% & $+2$\% \\ \hline
Mixed event    & $<+1$\% & $<+1$\% & $-4$\% & $+4$\% & $+20$\% & $+10$\% \\ \hline
Peri. reference    & $-1,+2$\% & $-2,+2$\% & $-2,+10$\% & $-10,+18$\% & $-10,+10$\% & $-2,+10$\% \\ \hline
Trig jet \pt    & - & $<+1$\% & - & $-6,+6$\% & - & $-10,+10$\% \\ \hline
Reject jet \pt    & - & $-2,+2$\% & - & $-5,+5$\% & - & $-10,+10$\% \\ \hline
Reject jet mult.    & - & $<+1$\% & - & $-5$\% & - & $-5$\% \\ \hline
Disabled HEC sector    & - & $<+1$\% & - & $ +10$\% & - & $ +20$\% \\ \hline
Jet-UE bias    & - & $ -5$\% & - & $ -15$\% & - & $ -25$\% \\ \hline
Total    & $-1,+2$\% & $-5,+4$\% & $-5,+10$\% & $ -20,+20$\% & $-10,+25$\% & $ -30,+25$\% \\ \hline
\end{tabular}
\caption{Systematic uncertainty summary for anisotropy coefficients $v_2$. The values are approximate, as they represent the average variation in each \pt range, and are reported relative to \vtwo. Negative and positive values indicate downward and upward uncertainties respectively. Commas separate the downward and upward uncertainty where applicable.
}
\label{tab:my_label}
\end{table}
 
\begin{table}[]
\centering
\begin{tabular}{|l|c|c|c|c|c|c|}
\hline
\multirow{2}{*}{Source} &  \multicolumn{2}{c|}{ $\pt < 2$~\GeV} & \multicolumn{2}{c|}{$\pt=2$--$10$~\GeV} & \multicolumn{2}{c|}{$\pt > 10$~\GeV} \\ \cline{2-7}
&  MBT & Jet & MBT & Jet & MBT & Jet \\ \hline
Track/trig Eff.    & $+1$\% & $+4$\% & $<+6$\% & $-7$\% & $-5$\% & $-7$\% \\ \hline
Mixed event    & $<+1$\% & $-4$\% & $-20$\% & $+6$\% & $-200$\% & $+20$\% \\ \hline
Peri. reference    & $-2,+2$\% & $-7,+10$\% & $-8,+8$\% & $-20,+10$\% & $-150,+100$\% & $-30,+30$\% \\ \hline
Trig jet \pt    & - & $<+1$\% & - & $-10,+10$\% & - & $-40,+40$\% \\ \hline
Reject jet \pt    & - & $-2,+2$\% & - & $-15,+15$\% & - & $-15,+15$\% \\ \hline
Reject jet mult.    & - & $<+1$\% & - & $+5$\% & - & $+20$\% \\ \hline
Disabled HEC sector    & - & $-2$\% & - & $ -5$\% & - & $ -10$\% \\ \hline
Total    & $-2,+2$\% & $-10,+10$\% & $-20,+10$\% & $ -30,+20$\% & $-250,+100$\% & $ -50,+50$\% \\ \hline
\end{tabular}
\caption{Systematic uncertainty summary for anisotropy coefficients $v_3$. The values are approximate, as they represent the average variation in each \pt range, and are reported relative to \vthree. Negative and positive values indicate downward and upward uncertainties respectively. Commas separate the downward and upward uncertainty where applicable.
}
\label{tab:v3}
\end{table}

The uncertainties associated with the measurement of particle pair yields are generated
from some of the variations discussed above, namely the track and trigger
efficiency variation, the trigger jet \pt threshold variation, and each B-particle
jet rejection variation. An additional variation was made to test the
assumption that the toward region contains all HS particles. The two transverse
region sides were tagged as having the minimum and maximum number of tracks out of the
two. The pair yields were, then, calculated using the minimum and maximum sides only,
as separate variations. This variation produces the dominant uncertainty in the relative
pair yields, defined in Eq.~(\ref{eq:pairfrac}), for all particle combinations at all \pt.

\section{Results}
\label{sec:result}
 
\begin{figure}[!htbp]
\begin{center}
\includegraphics[width=0.47\textwidth]{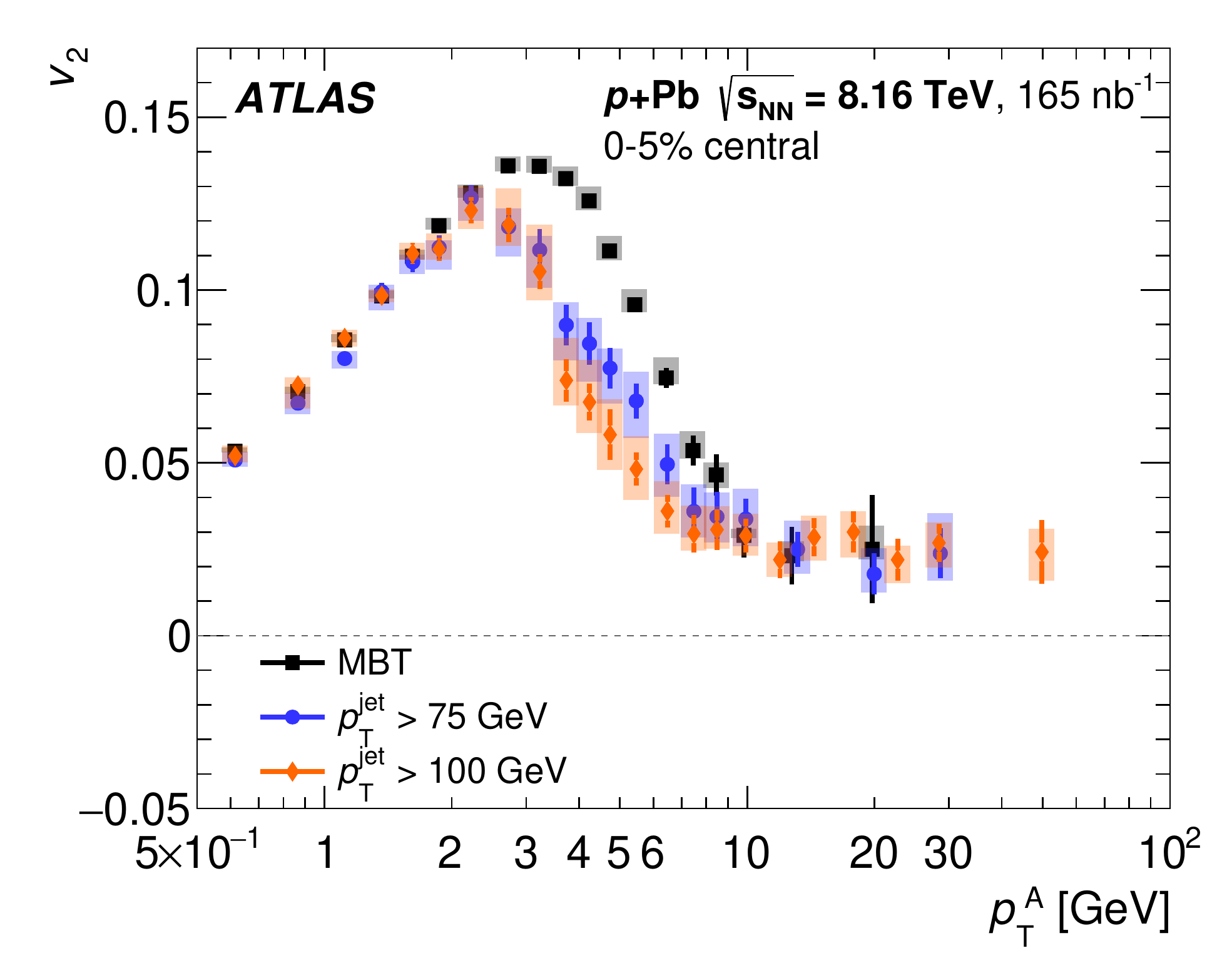}
\includegraphics[width=0.47\textwidth]{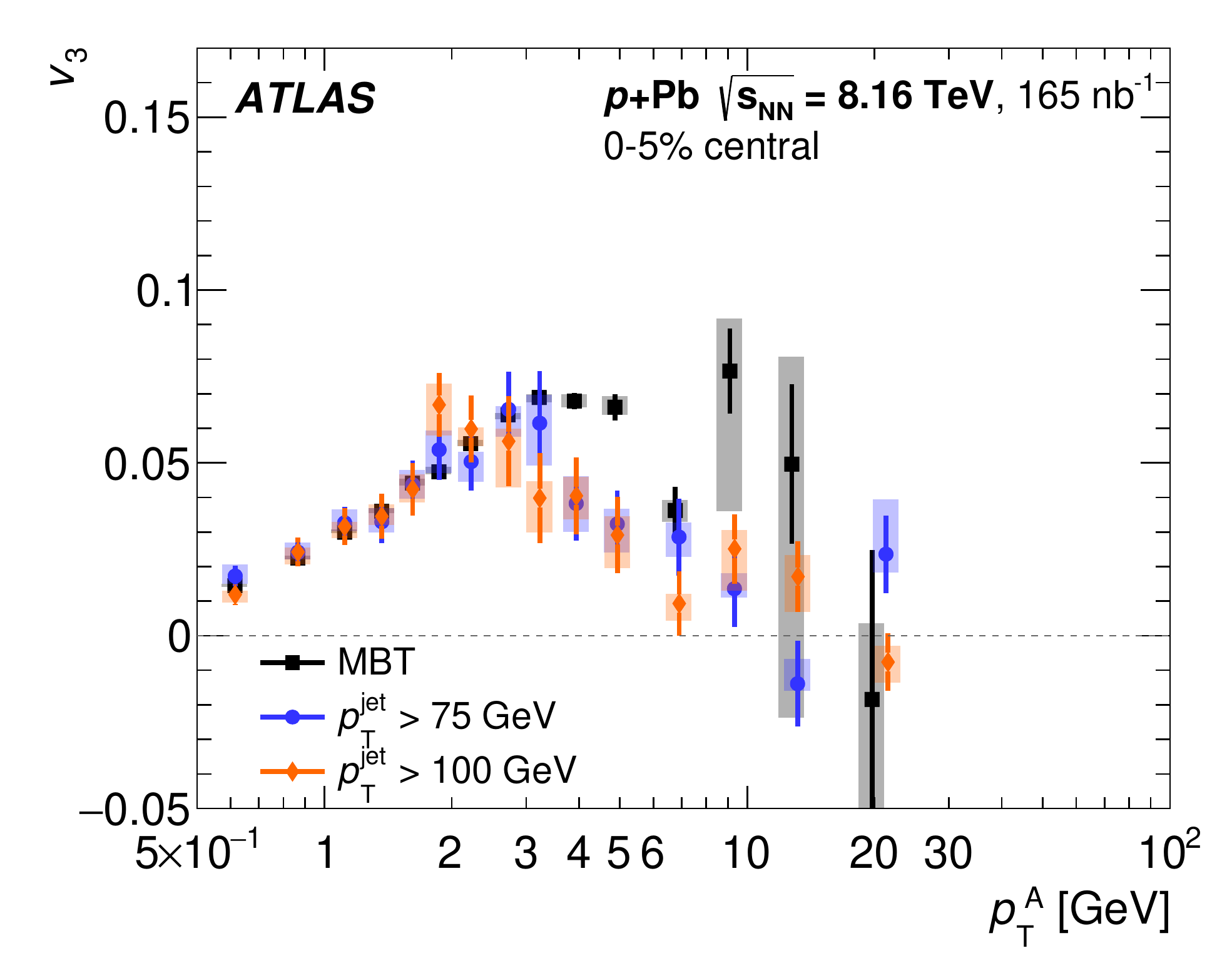}
\caption{Distribution of \vtwo{} (left) and \vthree{} (right) plotted as a function of the
A-particle \pt. Values from MBT events are plotted as black squares, and those
from events with jet $\pt>75$~\GeV{} and events with jet $\pt>100$~\GeV{} are
plotted as blue circles and orange diamonds respectively. Statistical uncertainties
are shown as narrow vertical lines on each point, and systematic uncertainties
are presented as coloured boxes behind the points.
\label{fig:res:v2v3}}
\end{center}
\end{figure}

\Fig{\ref{fig:res:v2v3}} shows the extracted second- (\vtwo{}) and third-order
(\vthree{}) anisotropy coefficients for the MBT events compared to those from both selections
of jet events plotted as a function of A-particle \pt in the range $0.5<\pt<100$~\GeV.
Each set of values is from events with the same 0--5\% centrality selection.
Points are located on the horizontal axis at the mean \pt of tracks within any given bin.
The \vtwo{} and \vthree{} coefficients increase as a function of \pt
in the low \pt region ($\pt <$ 2--3~\GeV), then decrease (2--3 $< \pt <$ 9~\GeV), and
finally plateau for high \pt ($\pt >$ ~9~\GeV{}). The \vtwo{} coefficients are
consistent with being independent of \pt for $\pt > 9$~\GeV{}, while the larger uncertainties
in the values of \vthree{} preclude any strong conclusion.
 
The \vtwo{} results show agreement within uncertainties between the MBT and jet events
for the low \pT ($\pt \lesssim 2$~\GeV{}) and high \pT ($\pt \gtrsim 9$~\GeV{})
regions. For the intermediate \pt region, the MBT events yield a higher \vtwo{} value
than jet events, although the trends are qualitatively similar. Similarly to \vtwo, the
\vthree{} results show agreement between the MBT and jet events for $\pt<2$~\GeV, and
higher values from MBT events for $\pt>2$~\GeV.
 
\begin{figure}[!htbp]
\begin{center}
\includegraphics[width=0.47\textwidth]{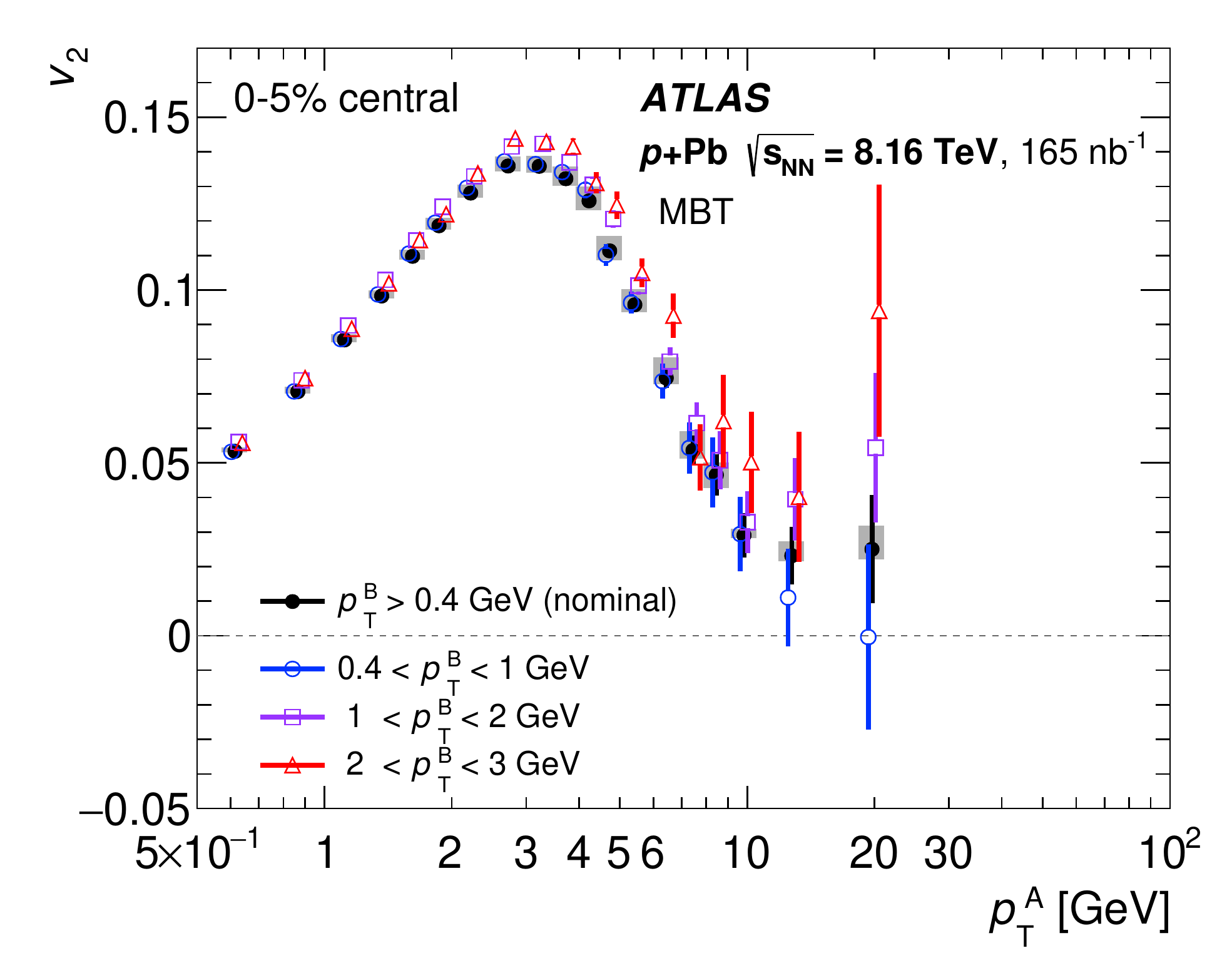}
 
\includegraphics[width=0.47\textwidth]{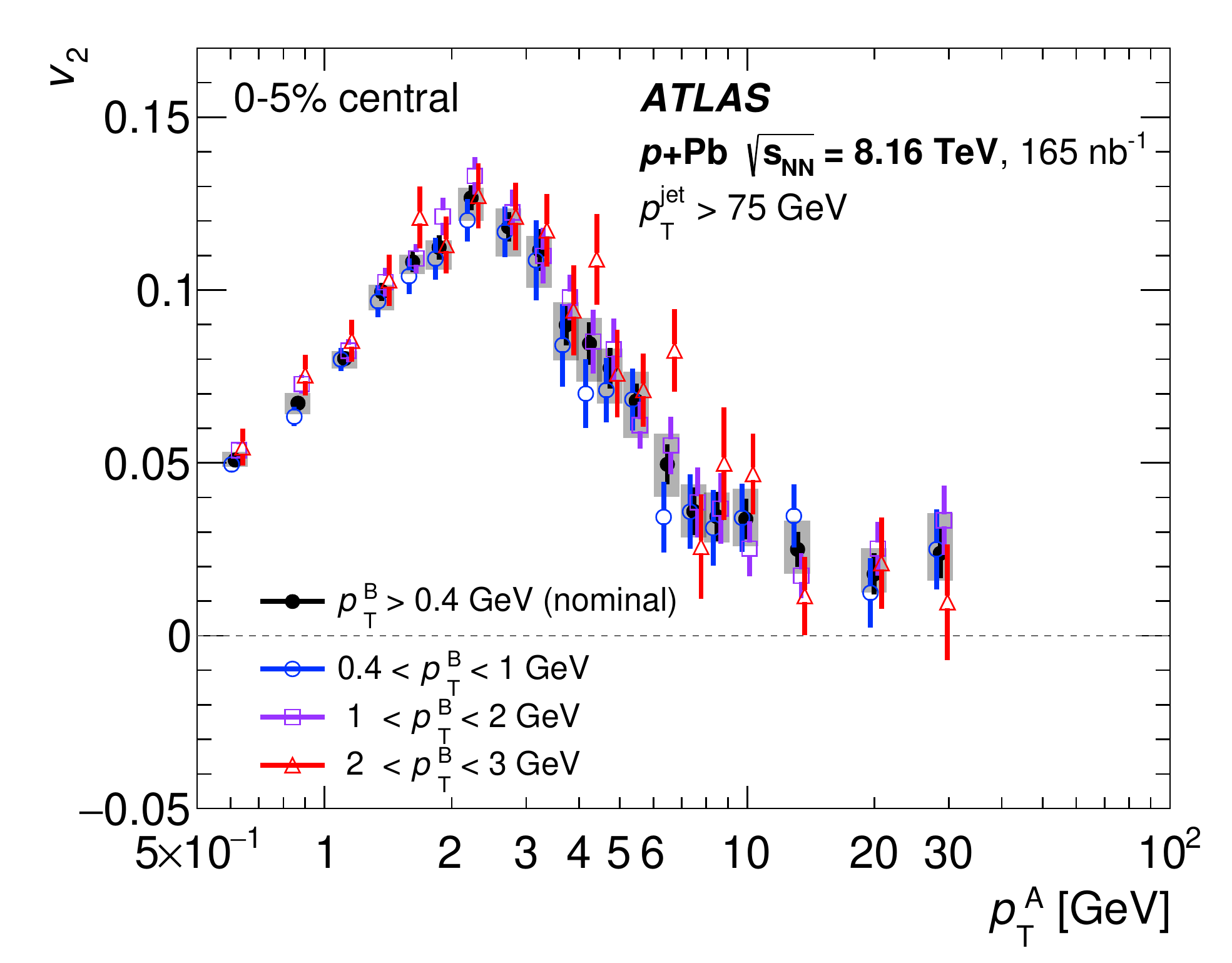}
\includegraphics[width=0.47\textwidth]{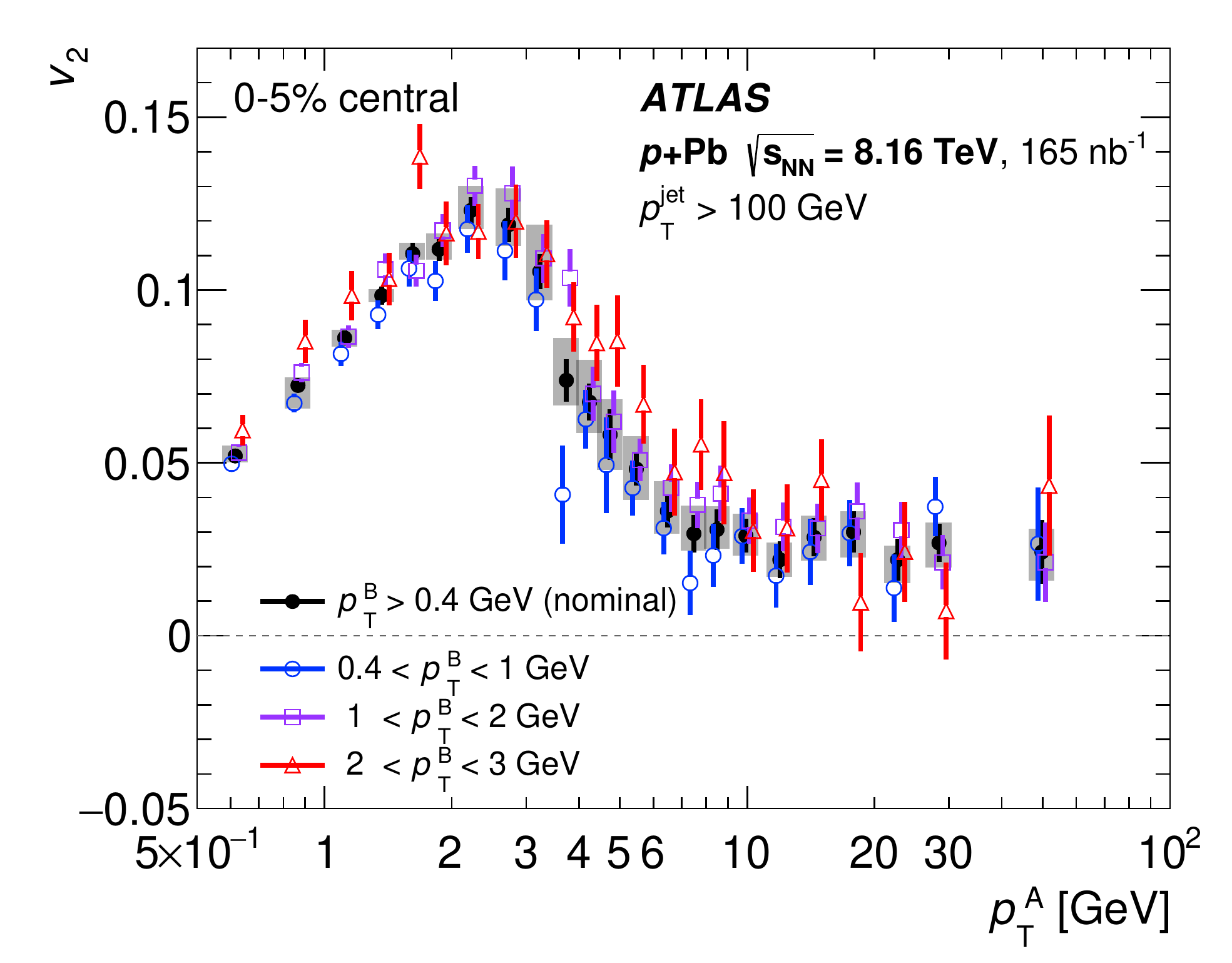}
\caption{Measured \vtwo{} values plotted as a function of the A-particle \pt for MBT events
(top), events with jet $\pt>75$~\GeV{} (bottom left), and events with jet $\pt>100$~\GeV{}
(bottom right). The nominal values (closed black circles) are overlaid with points generated
by making different B-particle \pt selections: $0.4<\pt^\mathrm{B}<1$~\GeV{} (blue open circles),
$1<\pt^\mathrm{B}<2$~\GeV{} (open violet squares), and $2<\pt^\mathrm{B}<3$~\GeV{} (open red triangles). The
points with different B-particle \pt selections are offset slightly from the nominal horizontal-axis
positions to make the uncertainties visible. For clarity, systematic uncertainties are omitted from
the three sets of restricted B-particle \pt ranges; they are, however, consistent with those from
the inclusive results and are highly correlated between the selections.
\label{fig:res:v2fac}}
\end{center}
\end{figure}
 
As mentioned in Section~\ref{sec:analysis}, if the measured anisotropy originates from a
global momentum field, the \vtwo{} and \vthree{} values, extracted for a given $\pt^\mathrm{A}$ range, will
be independent of B-particle selection. This assumption of factorisation is explicitly tested
by carrying out the analysis for different selections of $\pt^\mathrm{B}$. Figure~\ref{fig:res:v2fac}
shows the \vtwo{} values, from each event trigger, for the nominal results using
$\pt^\mathrm{B}>0.4$~\GeV{} overlaid with results using $0.4<\pt^\mathrm{B}<1$~\GeV{},
$1<\pt^\mathrm{B}<2$~\GeV{}, and $2<\pt^\mathrm{B}<3$~\GeV{}. The test shows factorisation
breaking at the level of 5\% for $\pt^\mathrm{A}<5$~\GeV{} in MBT events. However, at higher
$\pt^\mathrm{A}$, the differences grow with $\pt^\mathrm{A}$ to be 10--100\% from the nominal
values. For jet events, factorisation holds within about 10--20\% for all values of
$\pt^\mathrm{A}$, except for $4<\pt^\mathrm{A}<9$~\GeV{} in $\pt^\mathrm{jet}>100$~\GeV{} events,
where it is within about 30--40\%. Although the large uncertainties prevent strong conclusions
from being drawn, there is a hint of a difference in behaviour at high $\pt^\mathrm{A}$ where the
factorisation breaking is greater for MBT events than for jet events. This result could be due to the
B-particle jet rejection scheme used for the jet events. Correlations resulting from hard-process, e.g.\ from
back-to-back jets, specifically violate the factorisation assumption, and the B-particle jet
rejection dramatically limits the contribution from these processes from entering the
correlation functions in jet events. However, the correlations from MBT events have no such
rejection, and could, therefore, be more susceptible to hard-process correlations at high
$\pt^\mathrm{A}$.

\begin{figure}[!htbp]
\begin{center}
\includegraphics[width=0.47\textwidth]{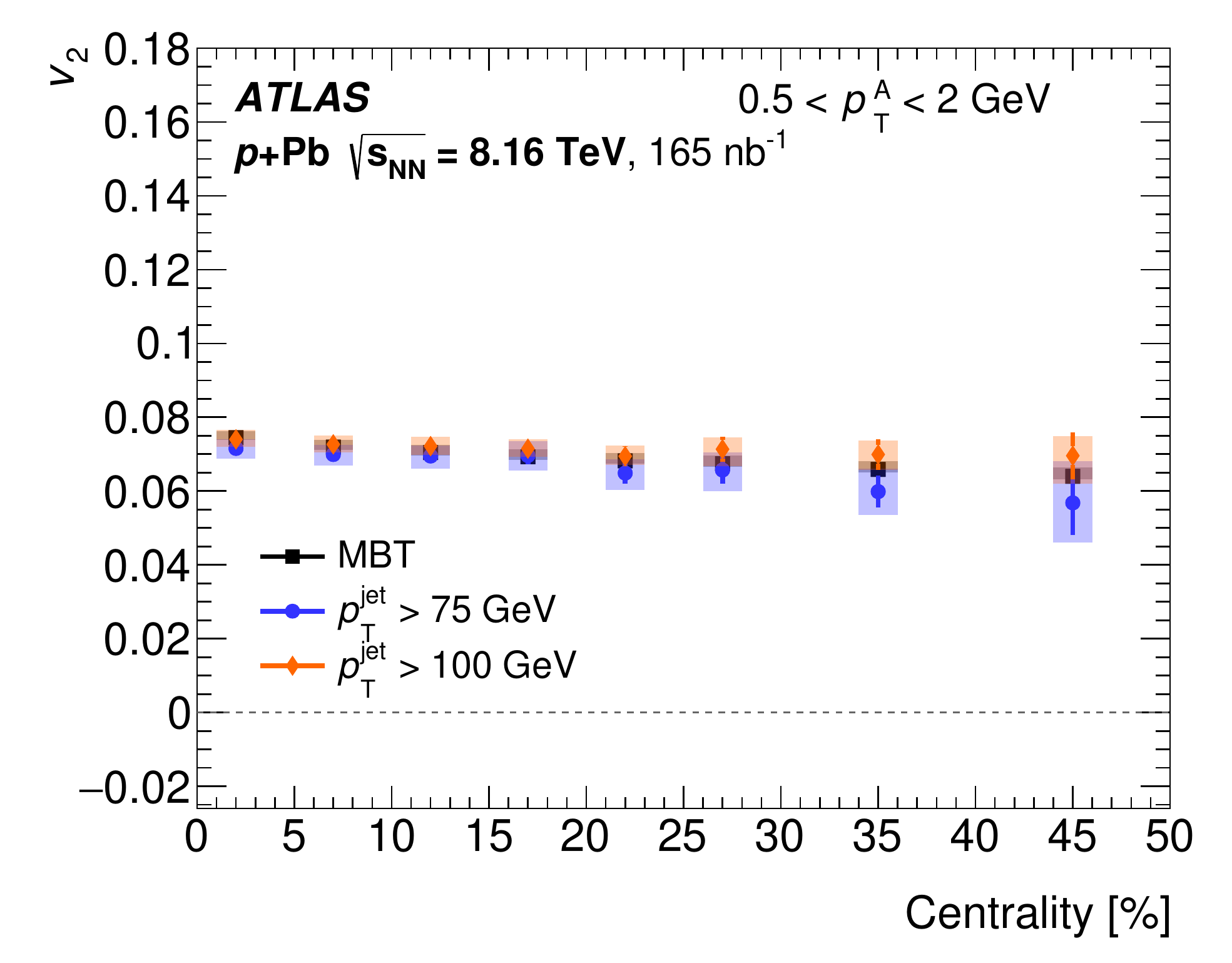}
\includegraphics[width=0.47\textwidth]{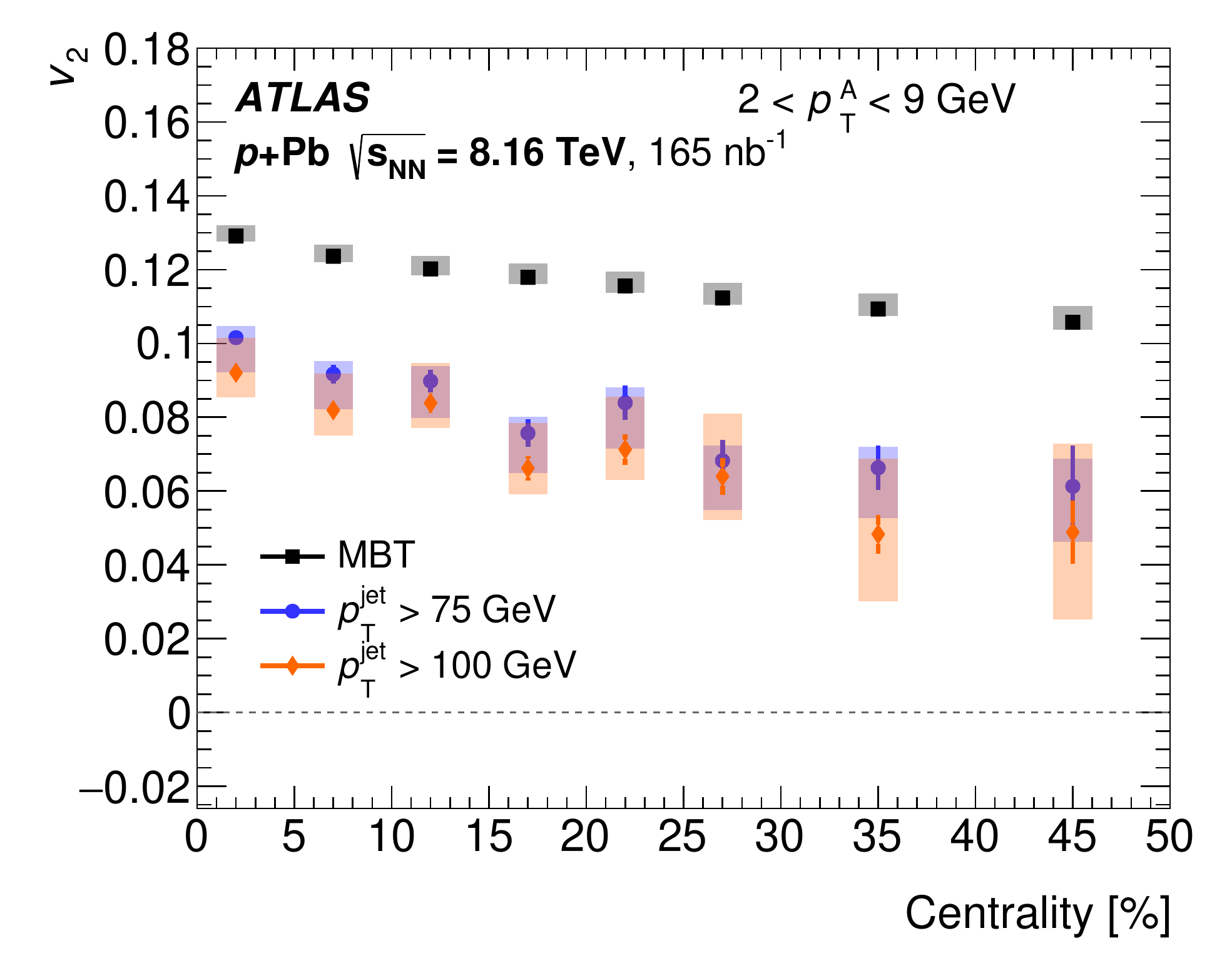}
\includegraphics[width=0.47\textwidth]{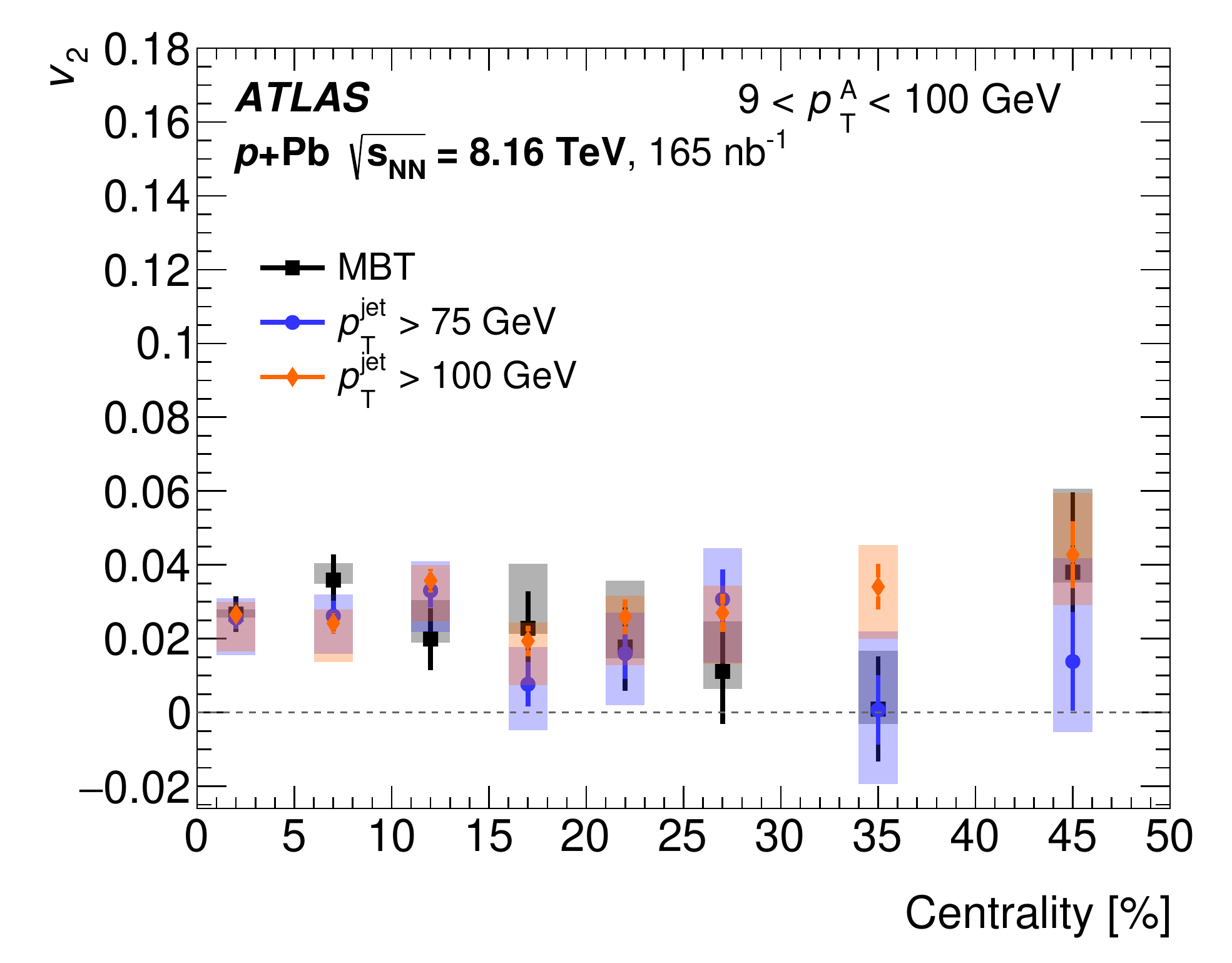}
\caption{Distribution of \vtwo{} plotted as a function of centrality for MBT events (black squares),
events with jet $\pt > 75$~\GeV{} (blue circles), and events with jet $\pt > 100$~\GeV{} (orange diamonds).
The results are obtained in three different selections of the A-particle \pt:
$0.5 < \pt < 2$~\GeV{} (top left), $2 < \pt < 9$~\GeV{} (top right), and $9 < \pt < 100$~\GeV{}
(bottom). Statistical uncertainties are shown as narrow vertical lines on each
point, and systematic uncertainties are presented as coloured boxes behind the points.}
\label{fig:res:v2cent}
\end{center}
\end{figure}
 
\Fig{\ref{fig:res:v2cent}} shows \vtwo{} plotted as a function of centrality for MBT events
and both classes of jet events. The results are
divided into three regions in A-particle \pt: $0.5 < \pt < 2$~\GeV{}, $2 < \pt < 9$~\GeV{},
and $9 < \pt < 100$~\GeV{}. The \vtwo{} results show agreement, within uncertainties,
between the MBT and jet events for \pt selections $0.5<\pt<2$~\GeV{} and $\pt>9$~\GeV{}
for all centralities and are found to be nearly independent of centrality. For
$2<\pt<9$~\GeV, the MBT events give a higher \vtwo{} value than the jet events, and all
three sets show a trend to lower values of \vtwo{} as the collisions become more peripheral.

\begin{figure}
\centering
\includegraphics[width=0.47\textwidth]{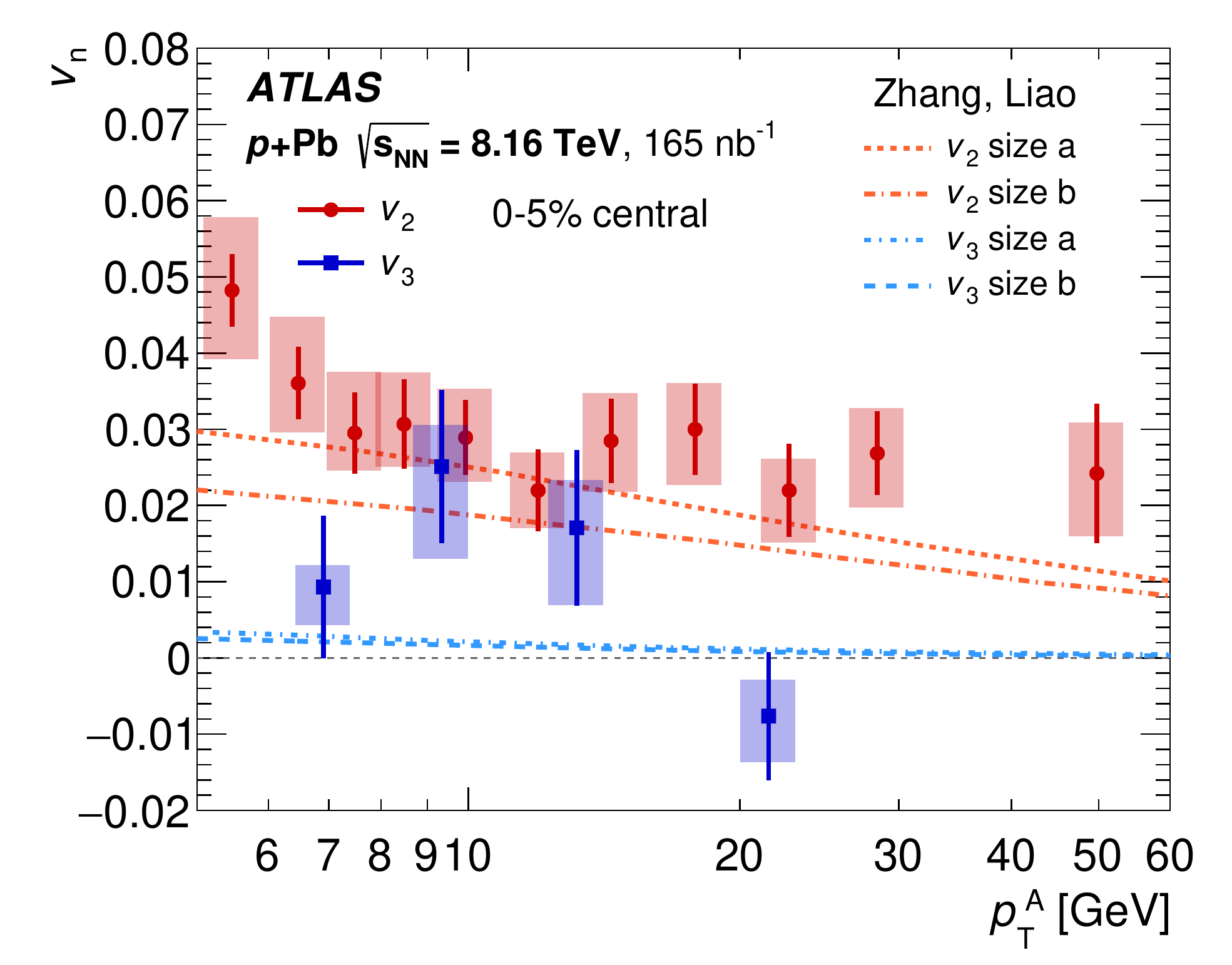}
\includegraphics[width=0.47\textwidth]{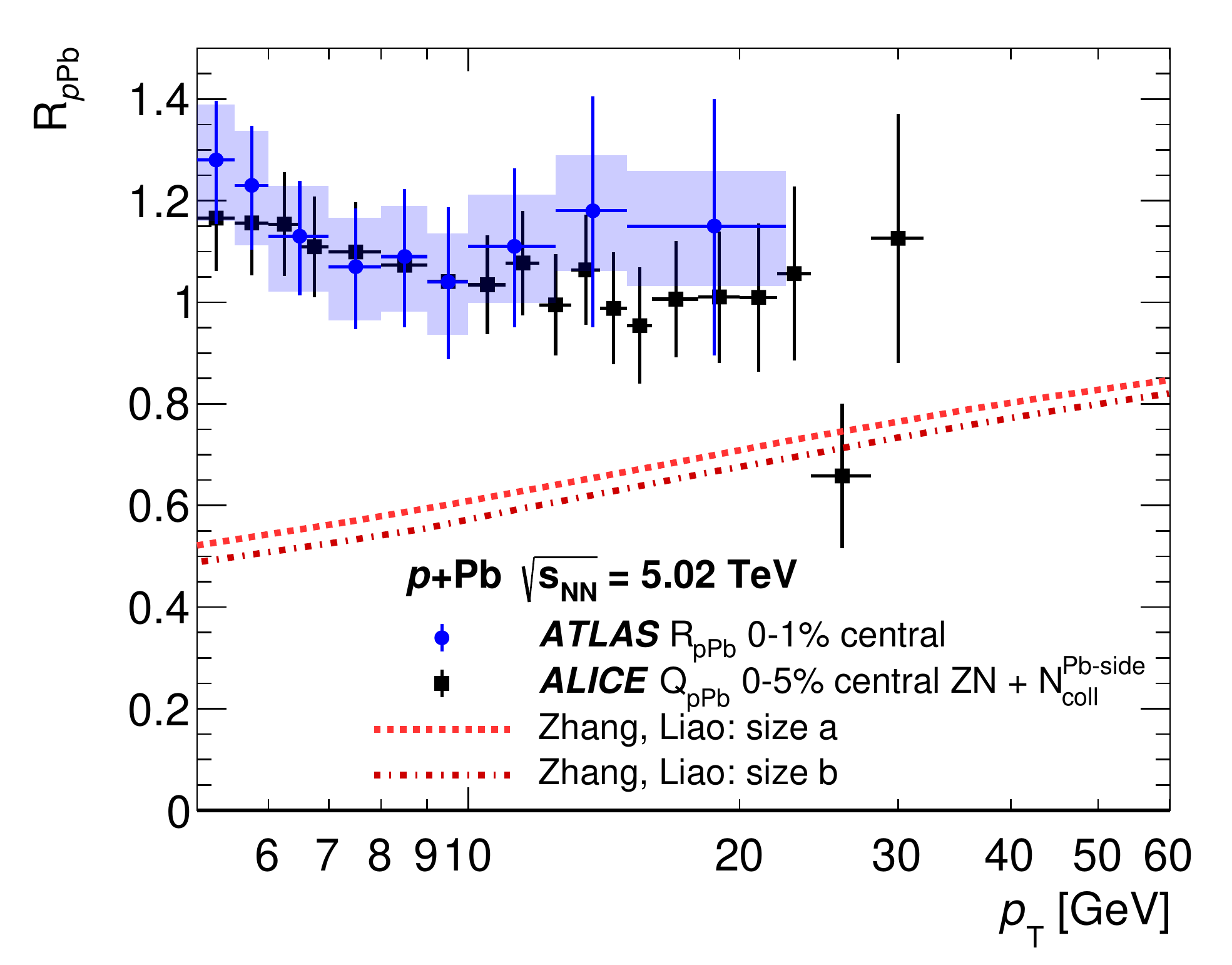}
\caption{Coefficients \vtwo{} and \vthree~ (left panel) and \RpPb (right panel) plotted as a function of particle \pt{} for \pPb{} collisions. The left panel is for central 0--5\% events from the jet $\pt > 100$~\GeV{} event sample. Statistical uncertainties are shown as narrow vertical lines on each point, and systematic uncertainties are presented as coloured boxes behind the points. The left panel has two sets of curves showing theoretical predictions from a jet quenching framework with two different initial geometries in 0--4\% central collisions~\cite{Zhang:2013oca}; the upper two (red/orange) are \vtwo{} for `size a' (dotted) and `size b' (dash-dotted) configurations, and the lower two (blue) are \vthree{} where the `size a' (dash-dotted) and `size b' (dashed) curves are nearly indistinguishable from each other. The right panel shows \RpPb data from ATLAS~\cite{HION-2012-14} and $Q_{p\mathrm{Pb}}$ data from ALICE~\cite{Adam:2014qja}. Theoretical calculations (red/orange lines) from \Ref{\cite{Zhang:2013oca}} are also shown in this panel; the dotted line gives the results of the `size a' configuration and the dash-dotted line gives the results of the `size b' configuration.
\label{fig:liao}}
\end{figure}

Focusing on the overall \pt dependence of the anisotropies, \Fig{\ref{fig:liao}}
(left panel) shows \vtwo{} and \vthree{} coefficients from events with jet
$\pT > 100$~\GeV{} compared with theoretical calculations from Ref.~\cite{Zhang:2013oca}.
This theoretical calculation, within the jet quenching paradigm, invokes a stronger parton
coupling to the QGP near the transition temperature, which helps to reduce the tension
in simultaneously matching the nucleus--nucleus high-\pT hadron spectrum suppression and the
azimuthal anisotropy \vtwo{}. The calculation tests two different initial \pPb geometries
referred to as `size a' and `size b', where the latter has a smaller
initial QGP volume. The predictions are slightly lower than the data for both \vtwo{} and
\vthree{}, and the `size a' curve is within two standard deviations of all points.
However, in the right panel of \Fig{\ref{fig:liao}}, the same calculation predicts a
substantial suppression of high-\pT hadrons, as expressed by the quantity $\RpPb =
\mathrm{d}^{2}N_{p\mathrm{Pb}}/\mathrm{d}\pt \mathrm{d}y / (\TpPb \times
\mathrm{d}^{2}\sigma_{pp}/\mathrm{d}\pt \mathrm{d}y)$ where \TpPb{} represents the
nuclear thickness of the Pb nucleus, as determined via a Monte Carlo Glauber calculation~\cite{Miller:2007ri}. Shown in comparison are published experimental results from ATLAS and ALICE for
\RpPb in central events that are consistent with no nuclear suppression, i.e.\
$\RpPb = 1$~\cite{Adam:2014qja,HION-2012-14}. The ALICE experiment uses the notation
$Q_{p\mathrm{Pb}}$ for the same quantity to describe a bias that may exist due to the
centrality categorisation.
There are uncertainties in the experimental measurements
related to the centrality or multiplicity selection in \pPb{} collisions, particularly in
determining the nuclear thickness value \TpPb{}.   However, there is
no indication of the large \RpPb{} suppression predicted by the jet quenching calculation.
Thus, the jet quenching calculation is disfavoured as it cannot simultaneously describe the
non-zero high-\pt azimuthal anisotropy and the lack of yield suppression.
 
\begin{figure}[!b]
\begin{center}
\includegraphics[width=0.57\textwidth]{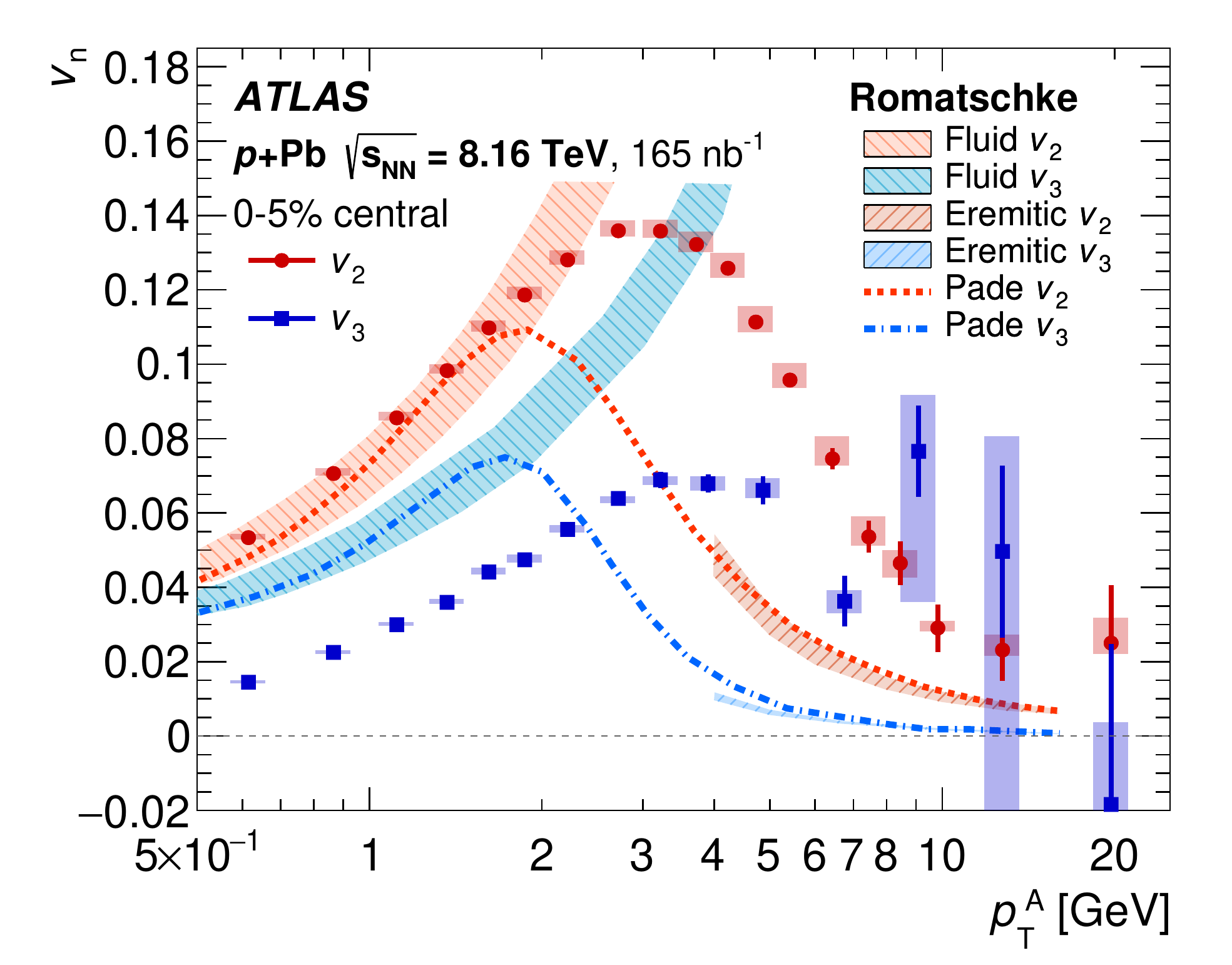}
\caption{Coefficients \vtwo{} and \vthree{} plotted as a function of \pt{} for central
0--5\% \pPb{} collisions from the MBT event sample. Theoretical calculations relevant
to the low-\pT{} regime from hydrodynamics and to the high-\pT{} regime from an
`eremitic' framework from Romatschke~\cite{Romatschke:2017ejr} are also shown. The lines
are Pad\'e-type fits connecting the two regimes, where the red dotted line is for \vtwo{} and
the blue dash-dotted line is for \vthree. Statistical uncertainties are shown
as narrow vertical lines on each point, and systematic uncertainties are presented as
coloured boxes behind the points.
\label{fig:res:Rom}}
\end{center}
\end{figure}

\Fig{\ref{fig:res:Rom}} shows the MBT \vtwo{} and \vthree{} coefficients compared with
theoretical calculations from Ref.~\cite{Romatschke:2017ejr}. The calculations
are derived from two opposite limits of kinetic theory. The low momentum
bands represent zeroth-order hydrodynamic calculations for high-multiplicity \pPb{}
events that give quantitative agreement with \vtwo{} up to $\pt = 2$~\GeV{} while
predicting values of \vthree{} that are too high.
Above some high \pT threshold, hadrons are expected to result, not from hydrodynamics,
but instead from jets where the resulting partons have the opposite limit than in
hydrodynamics, i.e.\ a large mean free path. To model this region, a non-hydrodynamic
`eremitic' expansion calculation (see Ref.~\cite{Romatschke:2017ejr} for the detailed
calculation), shown as the bands at high \pt, indicates slowly declining \vtwo{} and
\vthree{} coefficients. The dashed lines are a simple Pad\'e-type fit connecting
the two regimes~\cite{Romatschke:2017ejr}. The trends are qualitatively similar to those
in the data, although there is not quantitative agreement. In particular, the calculation
predicts values of \vtwo{} and \vthree{} substantially below the experimental results for
$\pt=$~4--15~\GeV{}. It should be noted that calculations presented in Ref.~\cite{Romatschke:2017ejr}
are performed, consistently between the hydrodynamic and eremitic components, only for
massless partons and with an ideal equation of state. Thus, one does not expect
quantitative agreement and is looking for rather qualitative trends. More
sophisticated treatments in the hydrodynamic regime result in better quantitative agreement
with the anisotropy coefficients at low \pt~\cite{Mantysaari:2017cni,Weller:2017tsr}.
It is worth highlighting that traditional parton energy-loss calculations connect the
high-\pT \vtwo{} with a suppression in the overall yield of high-\pT particles. The same
is true with this eremitic calculation, and thus, it should also be in contradistinction
to \pPb{} high-\pT experimental data indicating almost no suppression, i.e.\ jet
quenching.
 
\begin{figure}[!b]
\begin{center}
\includegraphics[width=0.44\textwidth]{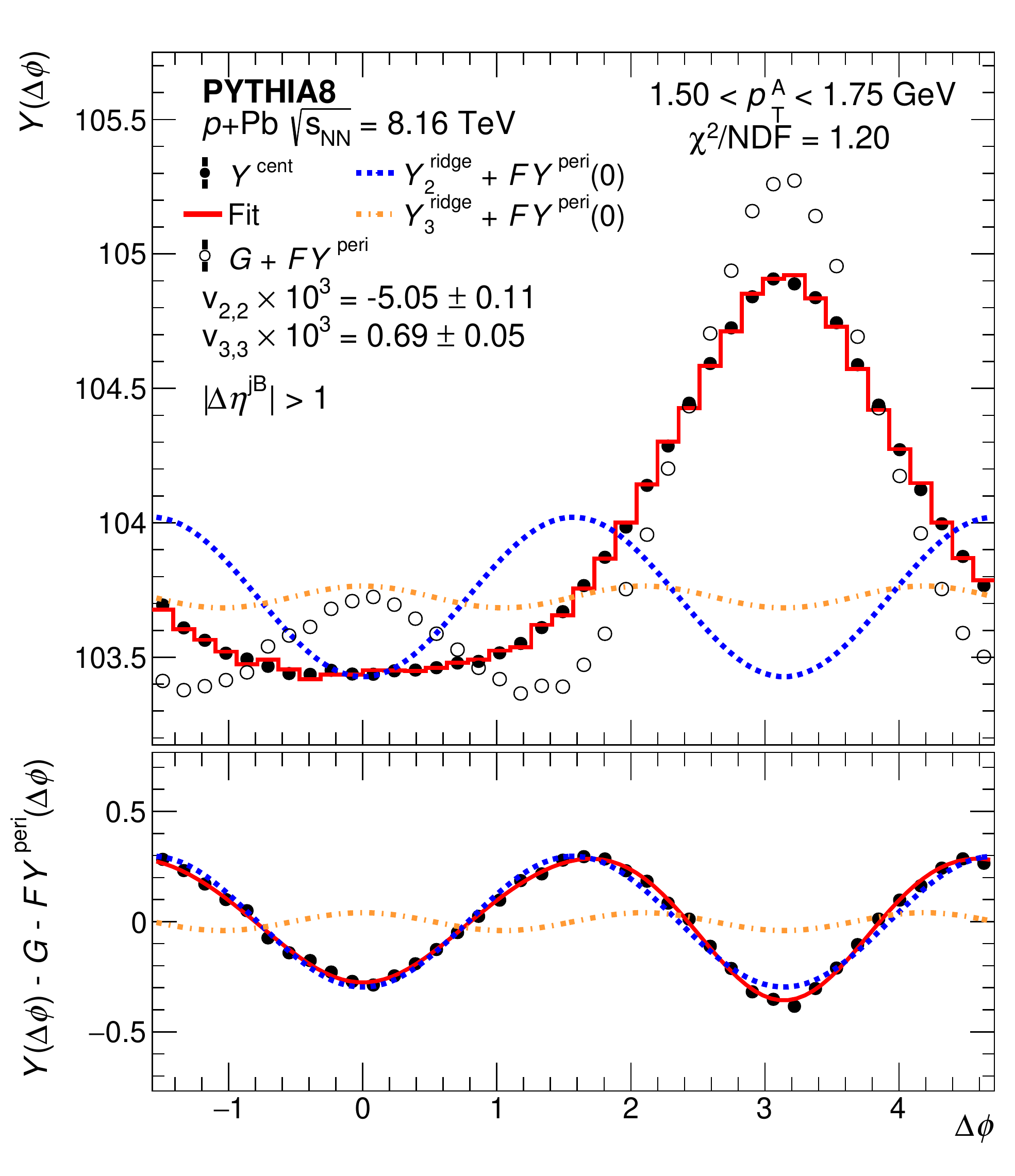}
\includegraphics[width=0.44\textwidth]{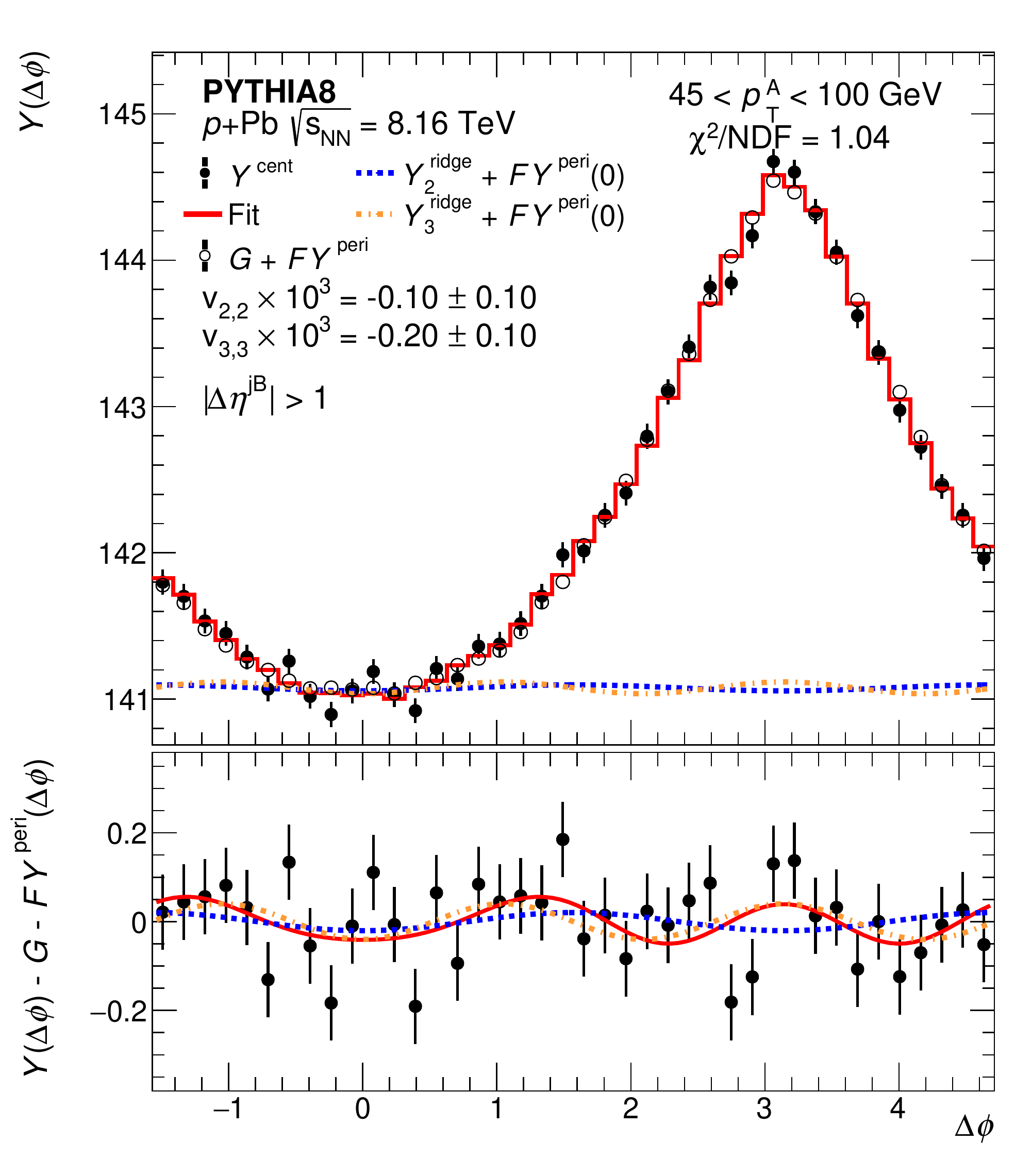}
\includegraphics[width=0.47\textwidth]{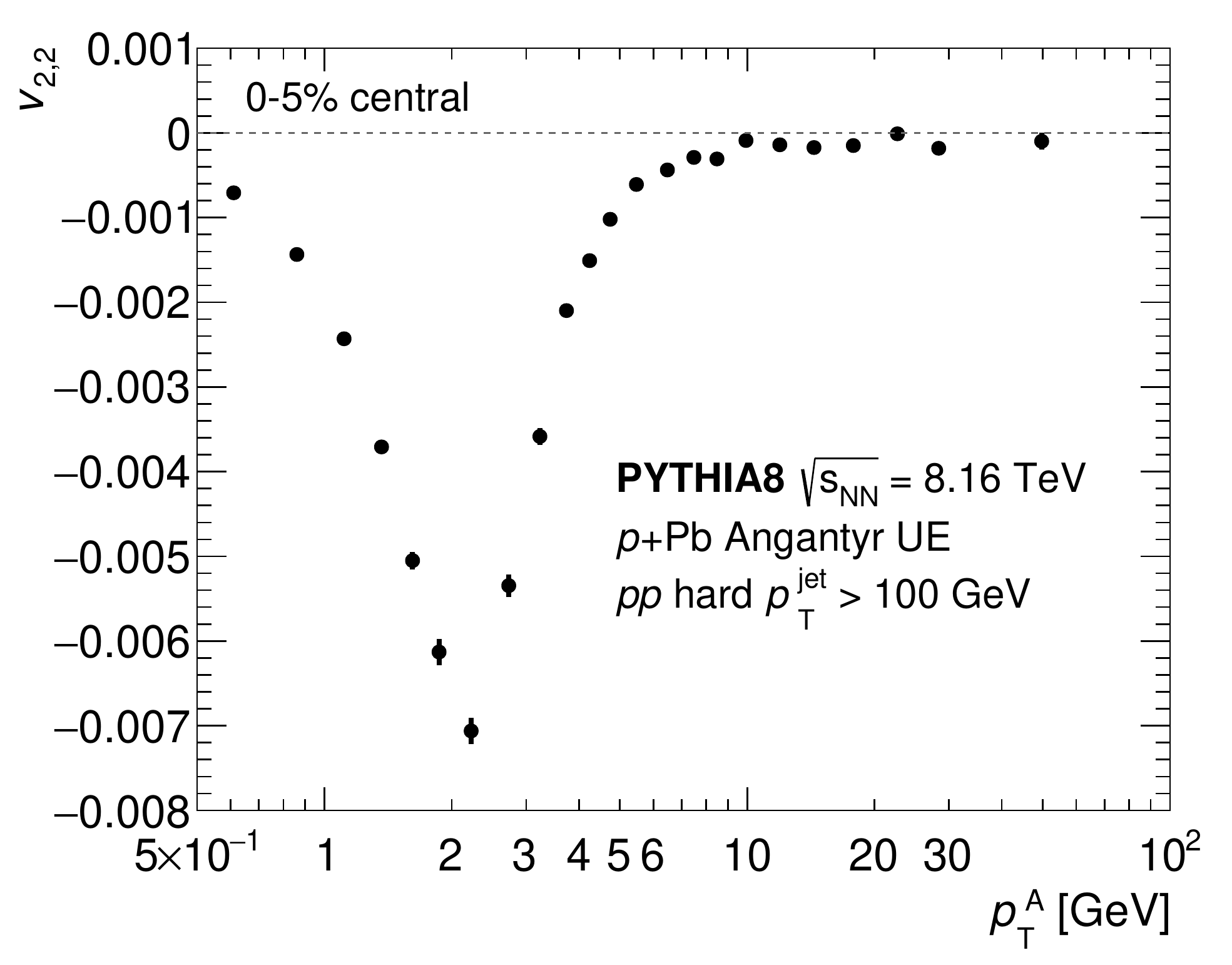}
\caption{Predictions of azimuthal anisotropy from \PYTHIAV{8} using the same two-particle formalism used for the data results. The events combine minimum-bias \pPb{} underlying events generated in the Angantyr framework with hard \pp{} events that require the presence of a jet with $\pt{} > 100$~\GeV{}. The two top plots show example correlation functions, with template fits, from a low particle-\pt{} selection (top left) and a high particle-\pt{} selection (top right). In the upper panels of the two top plots, the open circles show the scaled and shifted peripheral
template with uncertainties omitted, the closed circles show the central
data, and the red histogram shows the fit (template and harmonic functions). The
blue dashed line shows the second-order harmonic component, $Y_2^\mathrm{ridge}$, and the
orange dashed line shows the third-order harmonic component, $Y_3^\mathrm{ridge}$ (the $n=2$ and $n=3$
contributions to the sum in \Eqn{\ref{eqn:temp}}, respectively). The lower panels show the
difference between the central data and the peripheral template along with
the second and third harmonic functions. The resulting
$v_{2,2}$, $v_{3,3}$, and global fit $\chi^2$/NDF values are reported in the legends, where NDF~$=35$. The bottom
plot shows the extracted $v_{2,2}$ values as a function of A-particle \pt.
\label{fig:res:pythia}}
\end{center}
\end{figure}

Another possible source of the high-\pT{} anisotropies could lie in an initial-state effect,
potentially encoded in a model such as \PYTHIAV{8}. Shown in Figure~\ref{fig:res:pythia} is a
\PYTHIAV{8} calculation with hard\footnote{The term `hard' refers to \PYTHIAV{8} run with the
following settings: \texttt{HardQCD:all=on}, \texttt{PartonLevel:MPI=off}, and containing
a jet with $\pt> 100$~\GeV{}.} \pp events overlaid on minimum-bias \pPb{} events
generated in the default Angantyr framework~\cite{Bierlich:2018xfw}.
It is emphasised that this version of \PYTHIA does not include the recently developed string--string,
or so-called string shoving, implementation~\cite{BIERLICH2019499}. The generator-level
charged particles are then processed with the entire analysis procedure, including the non-flow
template fit. The result is a negative $v_{2,2}$ for all momenta, in contradistinction to the
experimental data. Further investigation reveals that \PYTHIAV{8} run in `hard' scattering mode has
correlations with large pseudorapidity separation between particle pairs as a result of the
specific implementation of initial-state radiation. This correlation is reduced in high-multiplicity events
because of the large number of uncorrelated UE particles, and thus results in a negative $v_{2,2}$
after subtracting the non-flow contribution.
 
\begin{figure}[!htbp]
\begin{center}
\includegraphics[width=0.57\textwidth]{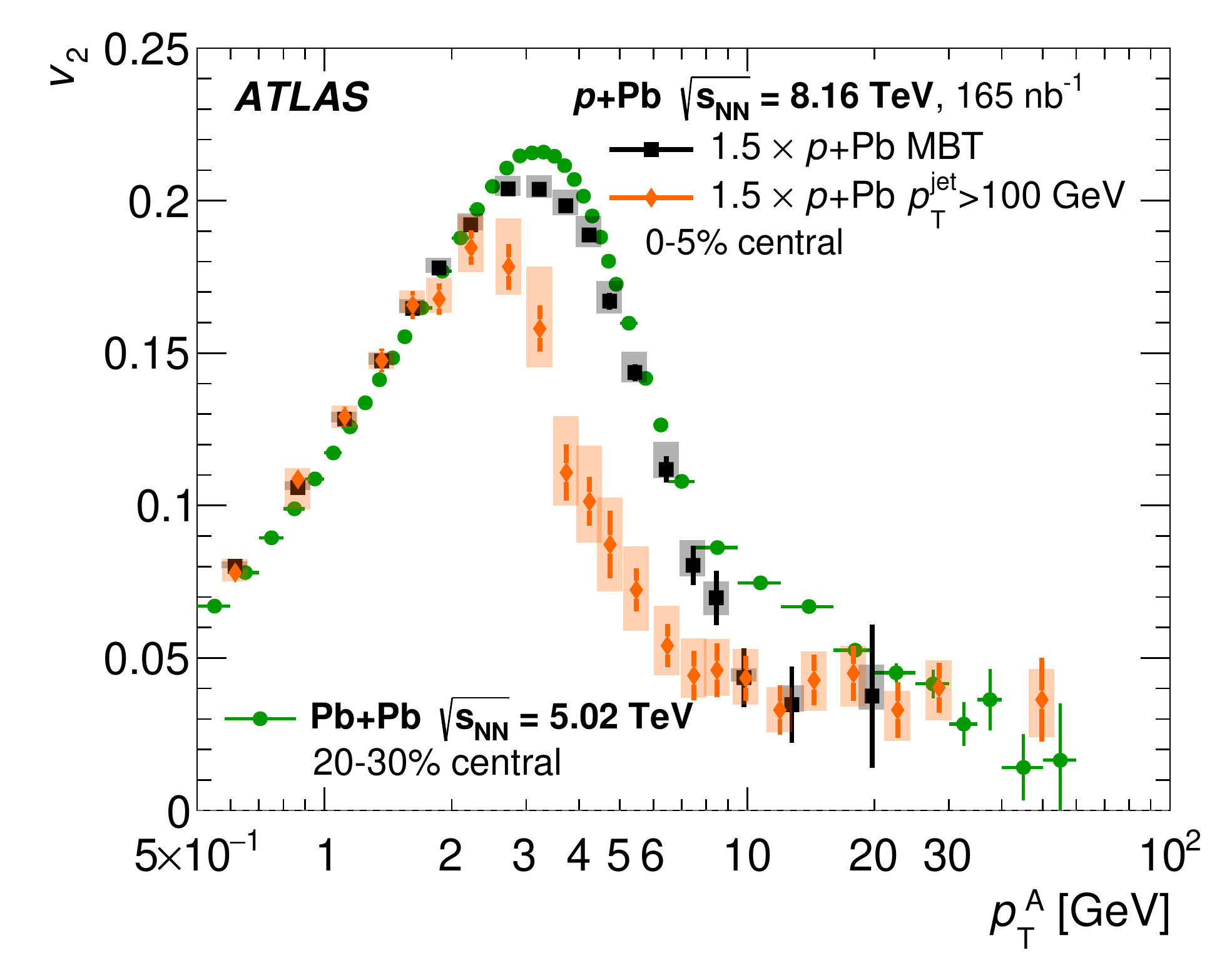}
\caption{Scaled \pPb{} \vtwo{} values plotted as a function of the A-particle \pt overlaid
with \vtwo{} from 20--30\% central \PbPb{} data at \sqn = 5.02~\TeV{}~\cite{HION-2016-06}.
Results from MBT and jet $\pt >100$~\GeV{} \pPb{} events are plotted as black squares and orange diamonds, respectively, and those from \PbPb{} are plotted as green circles.  Statistical uncertainties are shown as narrow vertical lines on each
point, and systematic uncertainties are presented as coloured boxes behind the points.
\label{fig:res:PbPbcomp}}
\end{center}
\end{figure}

\Fig{\ref{fig:res:PbPbcomp}} shows the published \PbPb{} results for
\vtwo{} as a function of \pt in the 20--30\% centrality
selection~\cite{HION-2016-06} compared to the \vtwo{} from both the MBT \pPb{} data and
\pPb{} containing a jet with $\pt{} > 100$~\GeV{}.
This \PbPb{} centrality range is selected because the spatial
elliptic eccentricity is approximately the same as in 0--5\% centrality \pPb{}
collisions~\cite{Loizides:2017ack}, despite having a much larger total particle
multiplicity. The overall trends for \PbPb{} \vtwo{} as a function of \pt are qualitatively similar to
those presented here for \pPb{} from MBT events and the jet events with jet $\pt > 100$~\GeV{}.
Both sets of the \pPb{} values are scaled by a single multiplicative factor (1.5) to match the \PbPb{} rise at low
\pt. After scaling, the MBT \pPb{} results quantitatively agree with those from the \PbPb system for $0.5 < \pt < 8$~\GeV,
except for a slight difference in the
peak value near $\pt \approx 3$~\GeV. For \pt above about 8~\GeV, the \PbPb{} results
indicate a slow decline of \vtwo{} values with increasing \pt, while the \pPb{} results
exhibit more of a plateau. Strikingly, the overall behaviour of the \vtwo{} values are quite similar.
 
As described above, the physics interpretations of the \PbPb elliptic
anisotropies are hydrodynamic flow at low \pt, differential jet quenching at high
\pt, and a transition between the two in the intermediate region of
approximately $2 < \pt < 10$~\GeV. Since these effects
all relate to the initial QGP geometric inhomogeneities, a common shape with a single
scaling factor for \pPb{} could indicate a common physics interpretation.
This scaling factor of 1.5, as empirically
determined, may be the result of slightly different initial spatial deformations, or
from the much larger \PbPb{} overall multiplicity, which enables a stronger translation
of spatial deformations into momentum space. For the high \pT region, this presents a
conundrum in that it is difficult for differential jet quenching to cause the \vtwo{}
anisotropy in \pPb{} collisions when there is no evidence for jet quenching overall.
These measurements showing non-zero high \pt{} \vtwo{} in \pPb{} collisions in the
absence the jet quenching observed in \PbPb{} collisions suggest there might be
additional contributions to \vtwo{} at high \pT in \PbPb{} collisions.
 
\begin{figure}[!htbp]
\begin{center}
\includegraphics[width=0.47\textwidth]{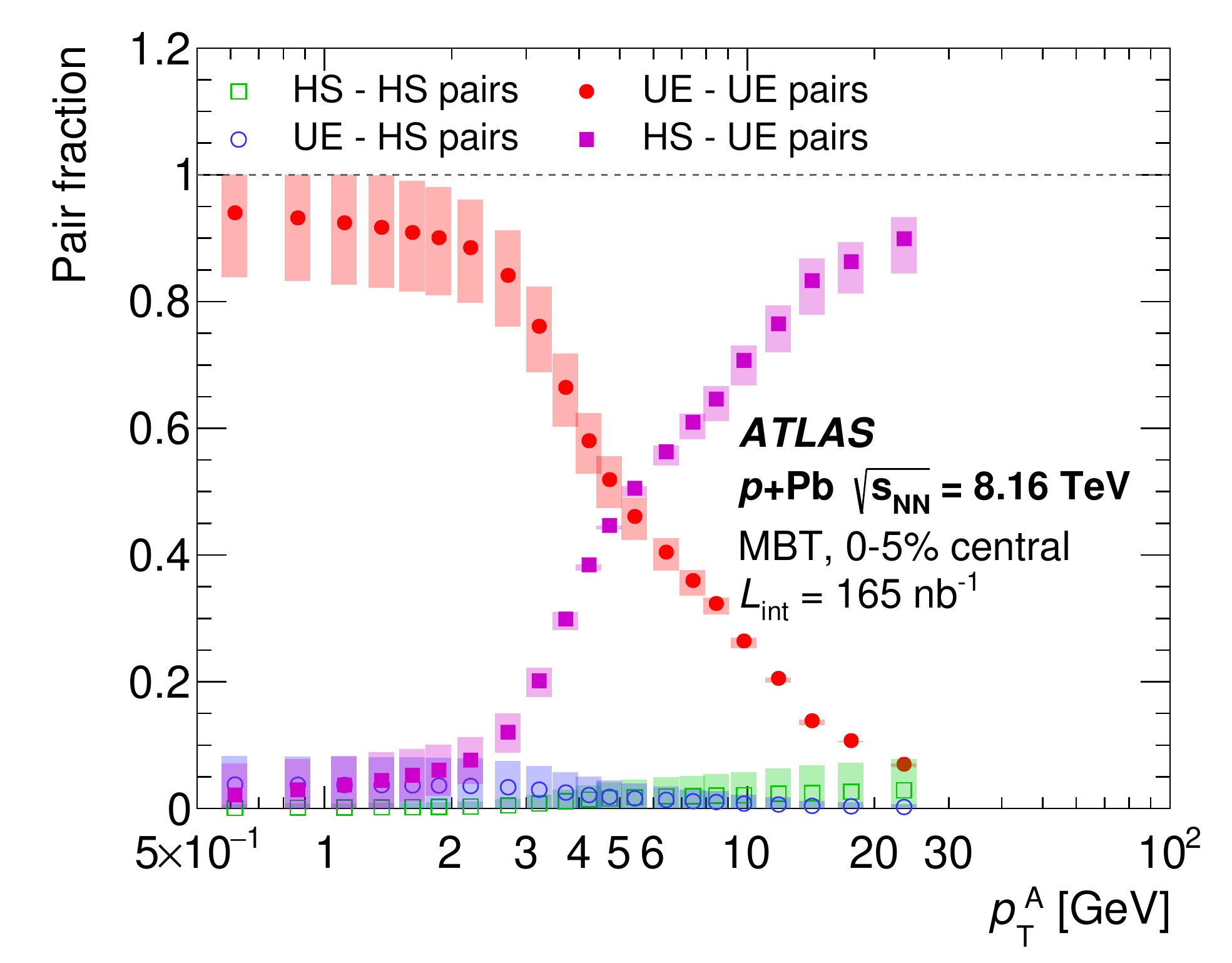}
 
\includegraphics[width=0.47\textwidth]{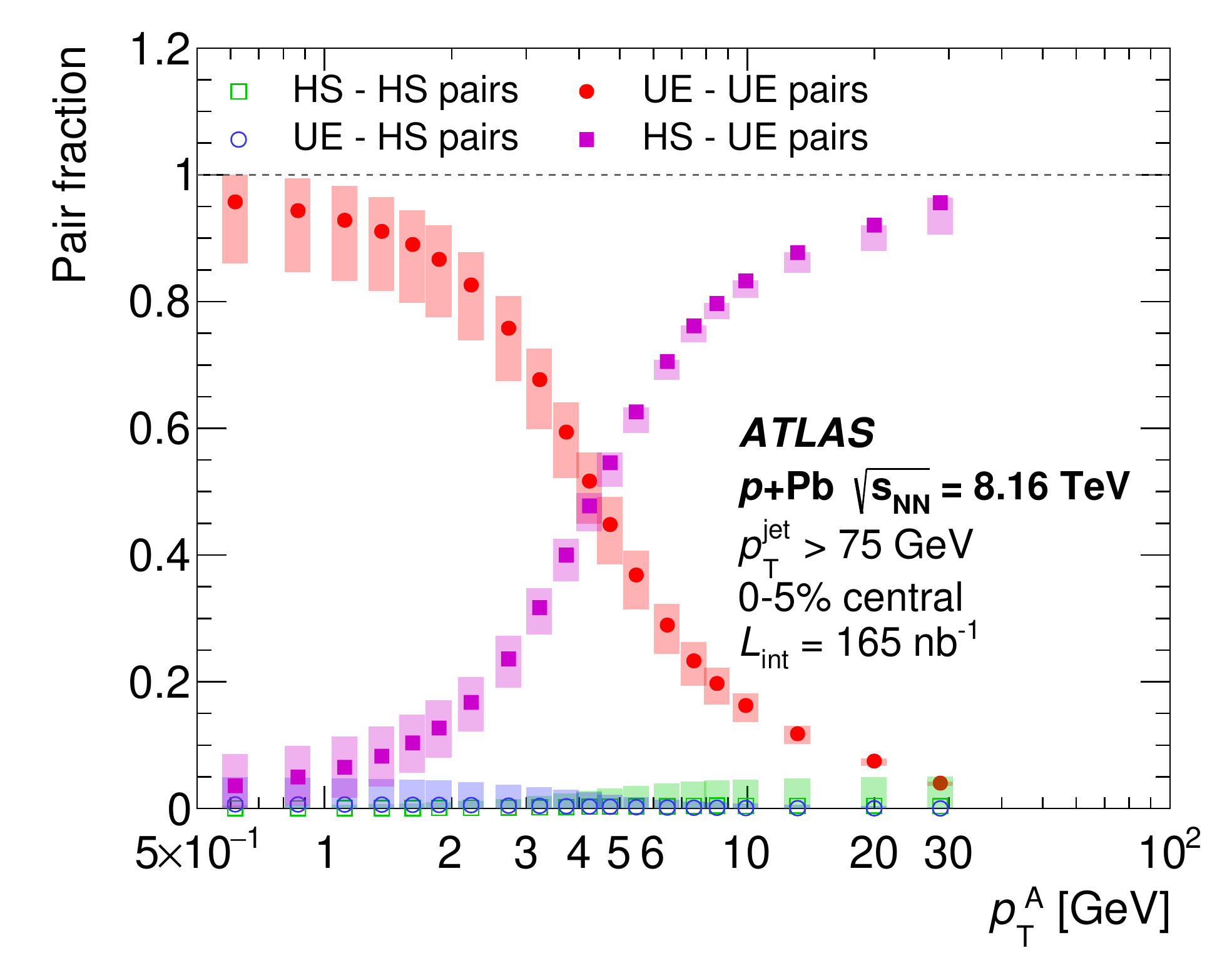}
\includegraphics[width=0.47\textwidth]{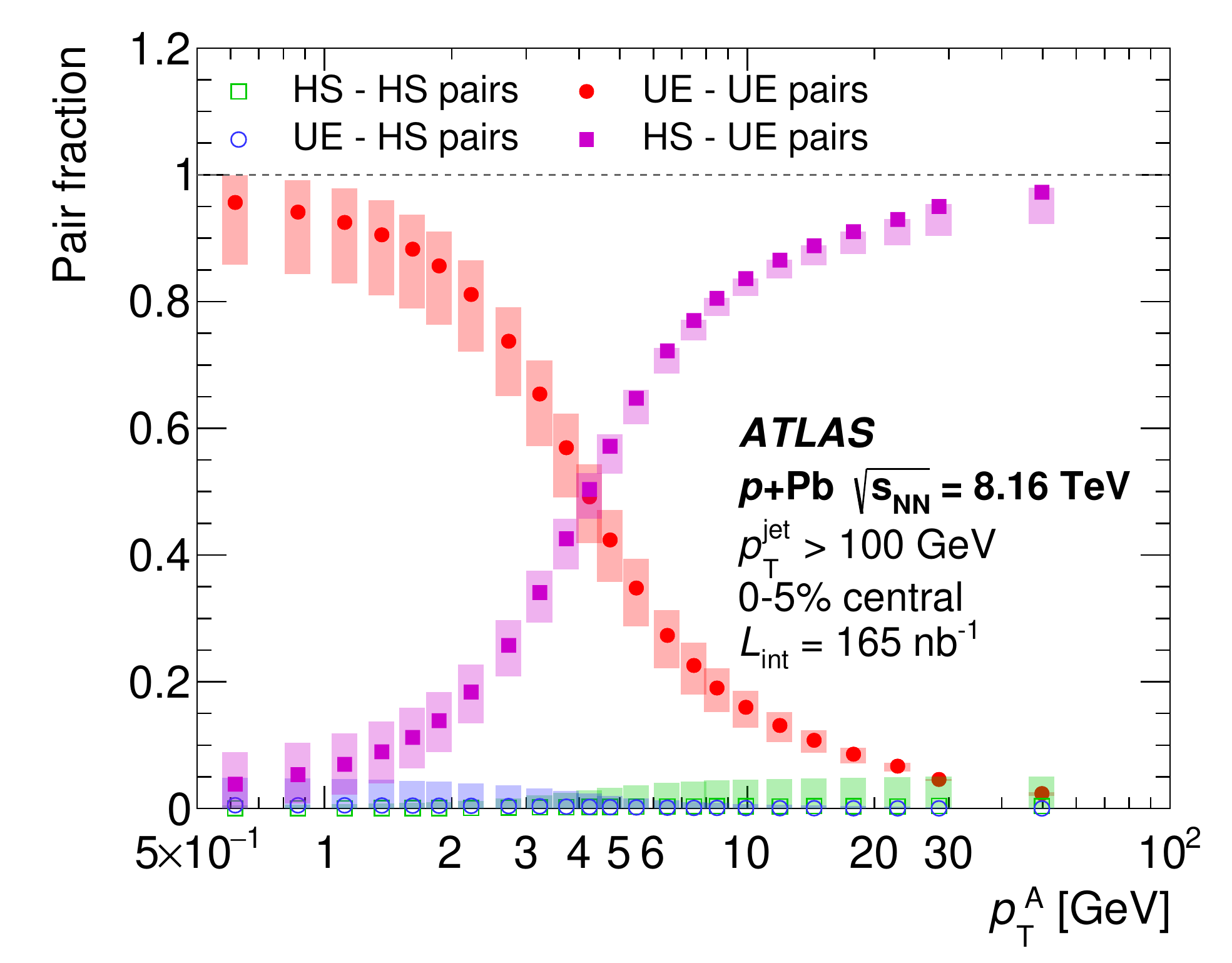}
\caption{Particle pair yield composition fractions for MBT events (top), events with jet
$\pt > 75$~\GeV{} (bottom left), and events with jet $\pt > 100$~\GeV{} (bottom right)
plotted as a function of the A-particle \pt.
Green and blue open circles represent HS--HS and UE--HS pairs, respectively, and
red and violet closed circles represent UE--UE and HS--UE pairs, respectively.
Statistical uncertainties are shown as narrow vertical lines on each
point, and systematic uncertainties are presented as coloured boxes behind the points.}
\label{fig:res:pairOverlay}
\end{center}
\end{figure}

\begin{figure}[!htbp]
\begin{center}
\includegraphics[width=0.57\textwidth]{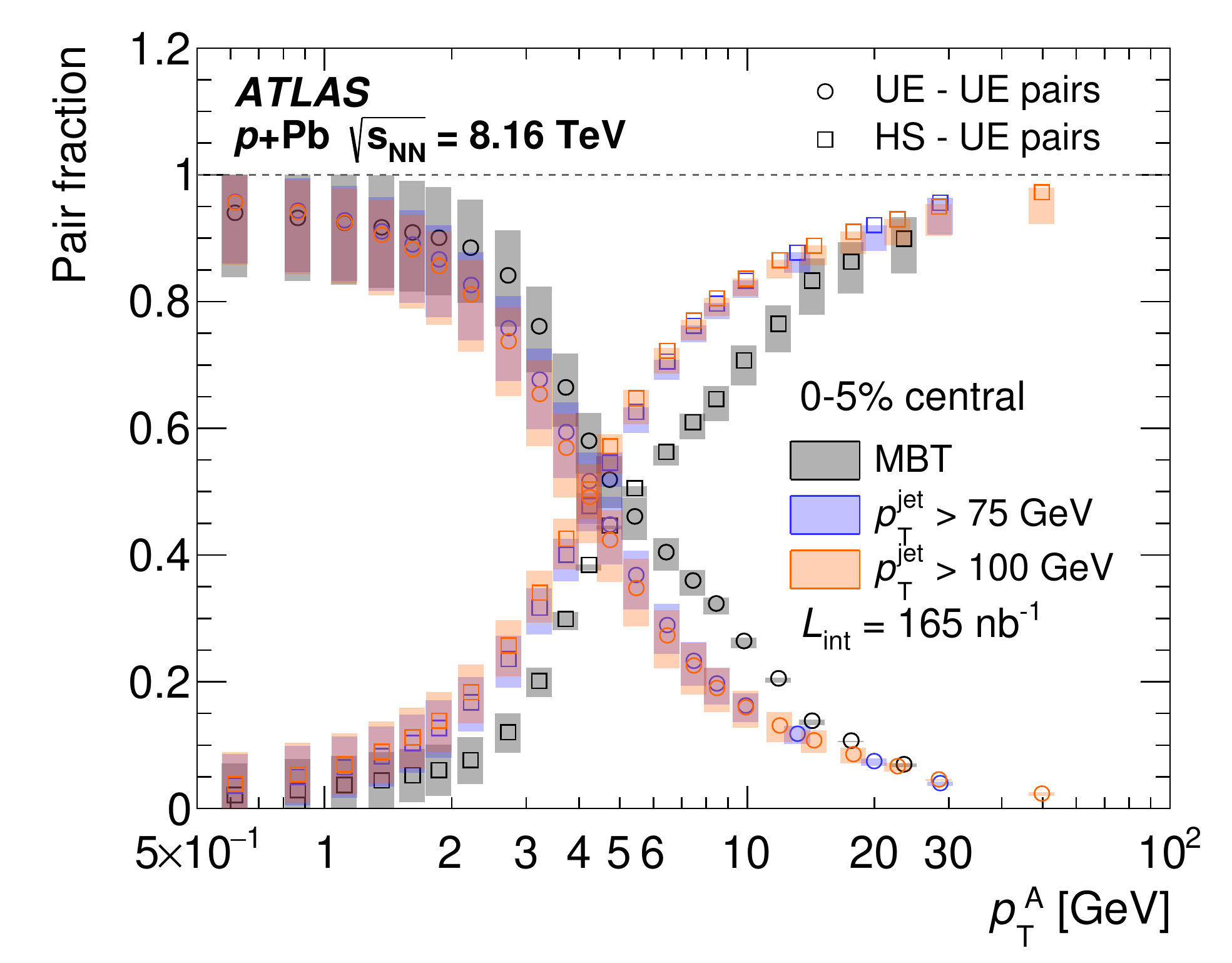}
\caption{Underlying event--underlying event (UE--UE) (open circles) and hard scatter--underlying
event (HS--UE) (open squares) particle-pair yield composition fractions for MBT events (black), events
with jet $\pt > 75$~\GeV{} (blue), and events with jet $\pt > 100$~\GeV{} (orange) plotted
as a function of the A-particle \pt. Statistical uncertainties are shown as narrow vertical
lines on each point, and systematic uncertainties are presented as coloured boxes behind the
points.}
\label{fig:res:pairMBjet}
\end{center}
\end{figure}

Returning to the issue of the difference in the intermediate \pt region between
the \pPb{} MBT and jet event results, the source of hadrons in this region
should be considered. As detailed previously, in a highly simplified picture one can classify
hadrons as originating from hard scatterings (HS) or from the underlying event (UE). Thus,
pairs of particles of A and B types can come from the combinations HS--HS, HS--UE, UE--HS, and
UE--UE. \Fig{\ref{fig:res:pairOverlay}} presents the measured pair fractions for both MBT and
jet, 0--5\% central events plotted as a function of the A-particle \pt. UE--UE pairs
dominate the correlation functions at low \pt in each case, and HS--UE
combinations dominate at high \pT. Combinations with HS B-particles are sub-dominant, because
there are fewer jet particles than UE particles in central events; for the
jet selected events, these combinations are further suppressed by the B-particle jet rejection
condition. \Fig{\ref{fig:res:pairMBjet}} shows the dominant contributions from the MBT and
jet events overlaid. Although the same qualitative behaviour is found in each case, the point
at which the HS--UE pairs become dominant over the other combinations is at a lower \pt for
jet events than for MBT events.
 
\begin{figure}[!htbp]
\begin{center}
\includegraphics[width=0.47\textwidth]{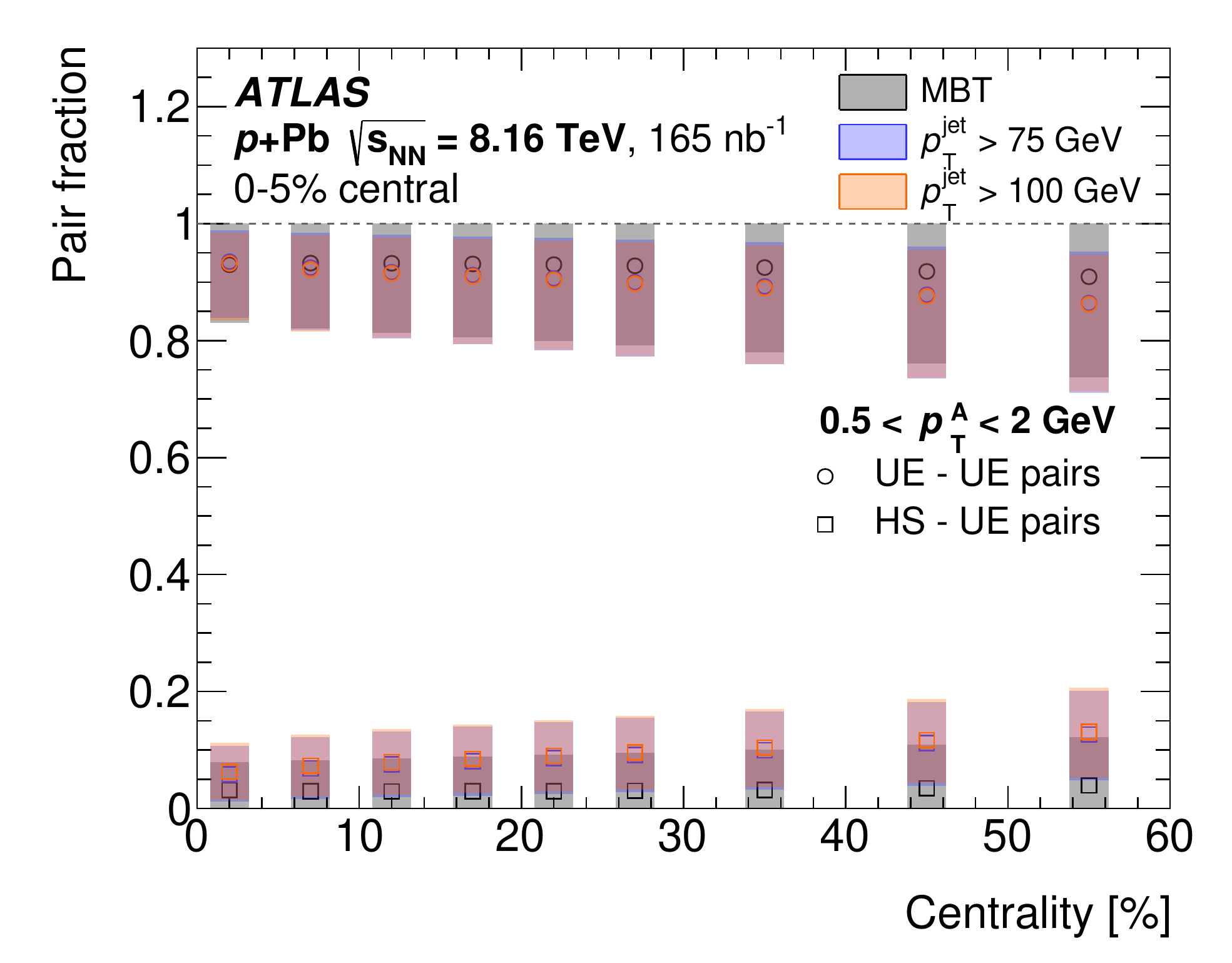}
\includegraphics[width=0.47\textwidth]{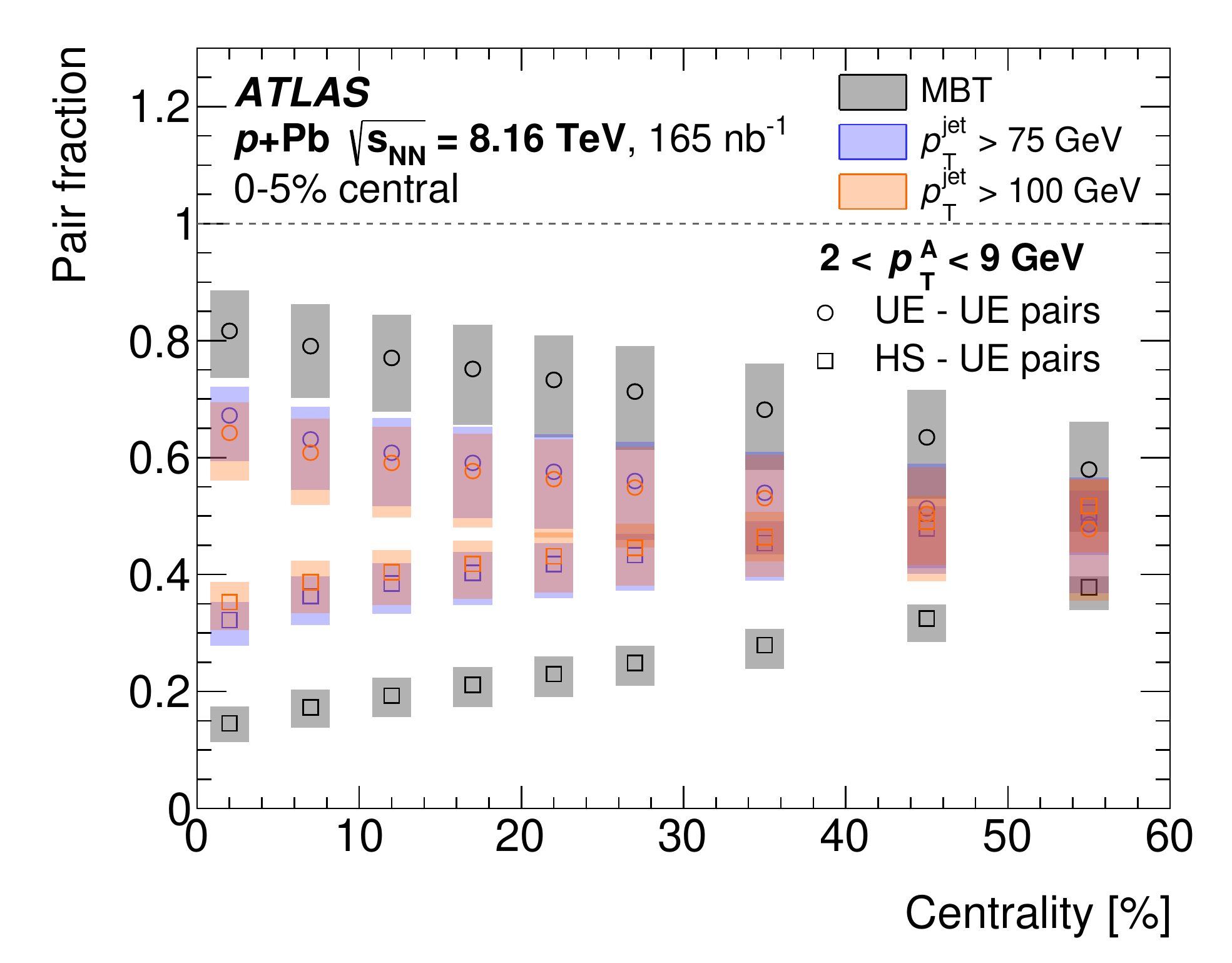}
\includegraphics[width=0.47\textwidth]{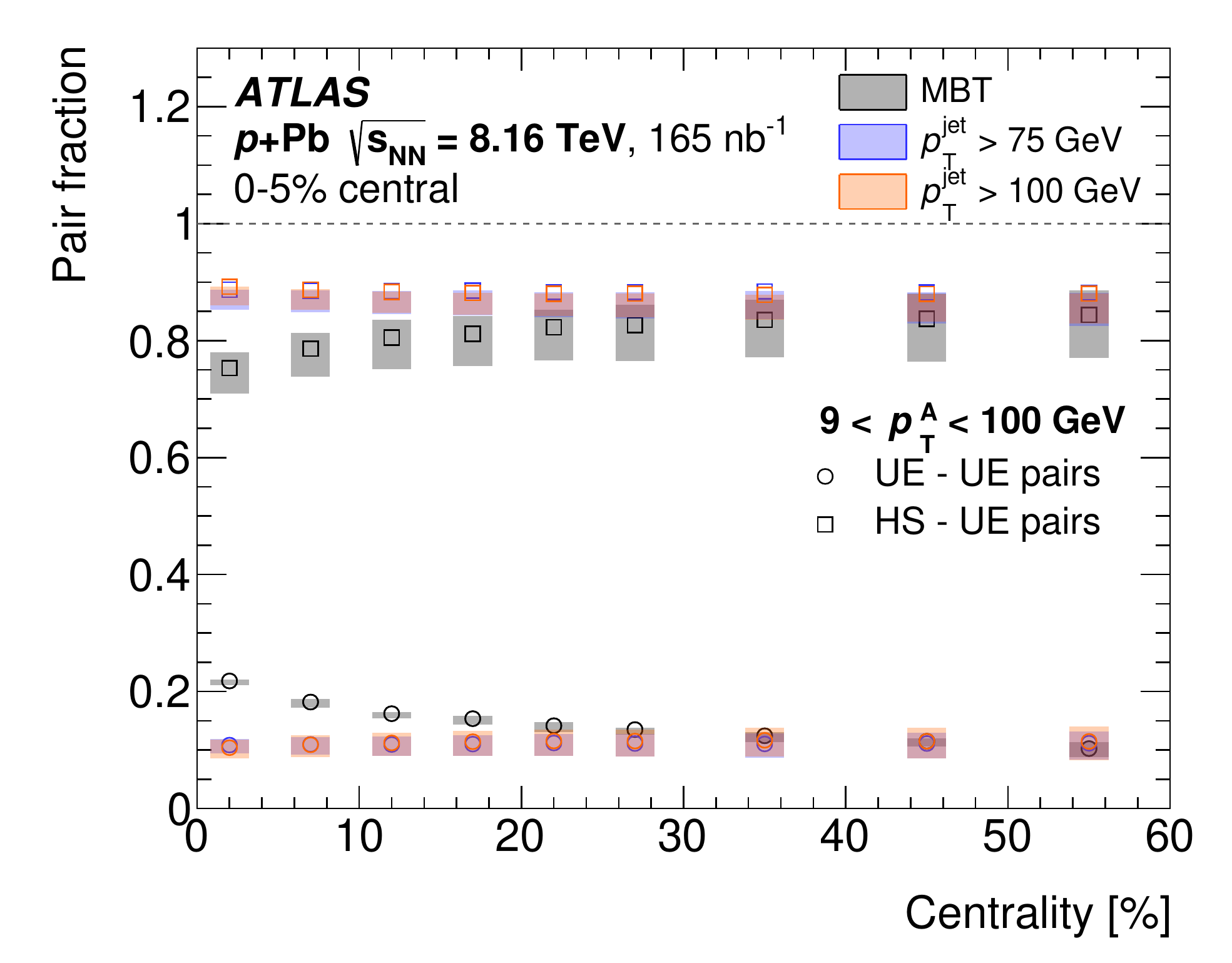}
\caption{Underlying event--underlying event (UE--UE) (open circles) and hard scatter--underlying
event (HS--UE) (open squares) particle-pair yield composition fractions for MBT events (black), events
with jet $\pt > 75$~\GeV{} (blue), and events with jet $\pt > 100$~\GeV{} (orange) plotted
as a function of event centrality. The results are obtained in three different selections of
the A-particle \pt: $0.5 < \pt < 2$~\GeV{} (top left), $2 < \pt < 9$~\GeV{} (top right), and $9 < \pt < 100$~\GeV{}
(bottom). Statistical uncertainties are shown as narrow vertical lines on each
point, and systematic uncertainties are presented as coloured boxes behind the points.}
\label{fig:res:pairMBjetCent}
\end{center}
\end{figure}

This behaviour can also be seen in \Fig{\ref{fig:res:pairMBjetCent}},
in which the pair fractions are plotted as a function of centrality, and again, the values
for MBT and jet events are overlaid. The centrality-dependent results are plotted for low,
medium, and high A-particle \pt ranges in the same way as in \Fig{\ref{fig:res:v2cent}}.
At low \pt, pair fractions from MBT and jet events agree, and in the mid-\pt transition
region, MBT events have a larger UE--UE contribution and smaller HS--UE contribution compared
to jet events. At high \pt, central events show a difference between UE--UE and HS--UE that
is reduced in more-peripheral events and absent for more peripheral than 25\% centrality. The overall trend
of the pair fractions with centrality is quite similar to that of \vtwo{} shown in
\Fig{\ref{fig:res:v2cent}}; little centrality dependence for low and high \pt, and significant
centrality dependence in addition to MBT--jet event ordering in the mid-\pt transition region.

Thus, a potential explanation for the lower \vtwo{} and \vthree{} in the intermediate
\pt region is simply that, in that region, the HS particles have lower anisotropy
coefficients than UE particles, and MBT events have a larger fraction of
UE--UE pairs than jet-triggered events. In the low and high \pT
regions, the same types of pairs dominate in both the MBT and jet-triggered events, namely
UE--UE and HS--UE respectively, and hence the anisotropy coefficients agree between the
event samples. If this explanation is correct, it also aids in understanding
Figure~\ref{fig:res:PbPbcomp} in which there is a significant difference between the
\pPb{} jet event \vtwo{} and the \PbPb{} \vtwo{} in the intermediate \pt region,
because the relative pair fractions are potentially significantly different.
 
This particle mixing picture is attractive in that it naturally explains the general
shape of the $\vtwo{}(\pt)$ and $\vthree{}(\pt)$ distributions as well as the ordering of
the different event samples. However, it is noted that the correspondence between the
differences in the flow coefficients and pair fractions is not quantitative; the
differences in the flow coefficients are fractionally much larger than the differences
in the pair fractions. Thus, there are either additional sources of correlation or
our assumptions are violated in some way (e.g.\ the two assumed HS and UE sources are
too simplistic or the measured pair fractions do not accurately represent the sources,
as is discussed in Section~\ref{sec:analysis}).
That said, for particle $\pt>20$~\GeV{}, where particle production in any model is
thought to arise mainly from jet fragmentation, the non-zero \vtwo{} demonstrates that
a positive correlation exists between hard (high \pt) and soft (low \pt) particles,
irrespective of the pair fractions.
 
\FloatBarrier

\section{Conclusion}
 
This paper presents Fourier coefficients of the azimuthal distribution of unidentified charged particles
from $165$~$\mathrm{nb}^{-1}$ of $\sqn = 8.16$~\TeV{} \pPb collisions at the LHC and
measured with the ATLAS detector. Results are presented separately for minimum-bias and
jet events, with jet \pt thresholds of 75 and 100~\GeV{}, as a function of particle \pt
and centrality. The results are extracted using two-particle azimuthal correlations combined with a non-flow
template fit procedure. 
The charged particle \pt{} dependence of \vtwo{} and \vthree{} is found assuming the factorisation of \vnn{}. The \vtwo{}
results are presented for charged-particle transverse momentum \pt~=~0.5--20~\GeV{} for minimum-bias events and \pt~=~0.5--50~\GeV{}
for jet-triggered events, and the \vthree{} results are for \pt~=~0.5--20~\GeV{} in both cases.
 
For charged particles with \pt{} between 0.5 and 2~\GeV{}, the $v_n$ results from each event selection
are quantitatively consistent with each other, rising steadily with \pt{}, and the \vtwo{}
coefficients are roughly independent of centrality. The \vtwo{} values at 0--5\% centrality
agree with those predicted by hydrodynamic calculations.
 
Between charged particle \pt of 2 and 9~\GeV{}, the $v_n$ values drop in each case, but are ordered with
minimum-bias events yielding the highest $v_n$ values and the jet events with jet $\pt > 100~\GeV$
the lowest. Charged particles
in this \pt{} range exhibit a significant centrality-dependent \vtwo, monotonically decreasing from
central to peripheral events. This behaviour can be qualitatively explained within a simplified
two-component model of particle production, in which the magnitude of the correlation in this
region is determined by the admixture of charged particles originating from soft and hard processes in the
given event selection. The measured particle pair yields support this qualitative argument.
 
For charged particles with \pt{} above 9~\GeV, the $v_n$ results are again consistent between MBT and jet
events. Although the uncertainties in the \vthree{} values make any quantitative statement
difficult, \vtwo{} plateaus at a value of 0.025 up to a \pt{} of 50~\GeV{}. This
result cannot be explained in the theoretical context of jet quenching or
eremitic expansion calculations while simultaneously describing
the observed lack of suppression of high-\pT{} hadron and jet yields in \pPb collisions.

\section*{Acknowledgements}
 

We thank CERN for the very successful operation of the LHC, as well as the
support staff from our institutions without whom ATLAS could not be
operated efficiently.
 
We acknowledge the support of ANPCyT, Argentina; YerPhI, Armenia; ARC, Australia; BMWFW and FWF, Austria; ANAS, Azerbaijan; SSTC, Belarus; CNPq and FAPESP, Brazil; NSERC, NRC and CFI, Canada; CERN; CONICYT, Chile; CAS, MOST and NSFC, China; COLCIENCIAS, Colombia; MSMT CR, MPO CR and VSC CR, Czech Republic; DNRF and DNSRC, Denmark; IN2P3-CNRS and CEA-DRF/IRFU, France; SRNSFG, Georgia; BMBF, HGF and MPG, Germany; GSRT, Greece; RGC and Hong Kong SAR, China; ISF and Benoziyo Center, Israel; INFN, Italy; MEXT and JSPS, Japan; CNRST, Morocco; NWO, Netherlands; RCN, Norway; MNiSW and NCN, Poland; FCT, Portugal; MNE/IFA, Romania; MES of Russia and NRC KI, Russia Federation; JINR; MESTD, Serbia; MSSR, Slovakia; ARRS and MIZ\v{S}, Slovenia; DST/NRF, South Africa; MINECO, Spain; SRC and Wallenberg Foundation, Sweden; SERI, SNSF and Cantons of Bern and Geneva, Switzerland; MOST, Taiwan; TAEK, Turkey; STFC, United Kingdom; DOE and NSF, United States of America. In addition, individual groups and members have received support from BCKDF, CANARIE, Compute Canada and CRC, Canada; ERC, ERDF, Horizon 2020, Marie Sk{\l}odowska-Curie Actions and COST, European Union; Investissements d'Avenir Labex, Investissements d'Avenir Idex and ANR, France; DFG and AvH Foundation, Germany; Herakleitos, Thales and Aristeia programmes co-financed by EU-ESF and the Greek NSRF, Greece; BSF-NSF and GIF, Israel; CERCA Programme Generalitat de Catalunya and PROMETEO Programme Generalitat Valenciana, Spain; G\"{o}ran Gustafssons Stiftelse, Sweden; The Royal Society and Leverhulme Trust, United Kingdom.
 
The crucial computing support from all WLCG partners is acknowledged gratefully, in particular from CERN, the ATLAS Tier-1 facilities at TRIUMF (Canada), NDGF (Denmark, Norway, Sweden), CC-IN2P3 (France), KIT/GridKA (Germany), INFN-CNAF (Italy), NL-T1 (Netherlands), PIC (Spain), ASGC (Taiwan), RAL (UK) and BNL (USA), the Tier-2 facilities worldwide and large non-WLCG resource providers. Major contributors of computing resources are listed in Ref.~\cite{ATL-GEN-PUB-2016-002}.
 

\printbibliography
 
\clearpage
 
\begin{flushleft}
{\Large The ATLAS Collaboration}

\bigskip

G.~Aad$^\textrm{\scriptsize 101}$,    
B.~Abbott$^\textrm{\scriptsize 128}$,    
D.C.~Abbott$^\textrm{\scriptsize 102}$,    
A.~Abed~Abud$^\textrm{\scriptsize 36}$,    
K.~Abeling$^\textrm{\scriptsize 53}$,    
D.K.~Abhayasinghe$^\textrm{\scriptsize 93}$,    
S.H.~Abidi$^\textrm{\scriptsize 167}$,    
O.S.~AbouZeid$^\textrm{\scriptsize 40}$,    
N.L.~Abraham$^\textrm{\scriptsize 156}$,    
H.~Abramowicz$^\textrm{\scriptsize 161}$
% ...
\bigskip
\\

% ...
$^{36}$CERN, Geneva; Switzerland.\\
% ...
$^{40}$Niels Bohr Institute, University of Copenhagen, Copenhagen; Denmark.\\
% ...
$^{53}$II. Physikalisches Institut, Georg-August-Universit\"{a}t G\"ottingen, G\"ottingen; Germany.\\
% ...
$^{93}$Department of Physics, Royal Holloway University of London, Egham; United Kingdom.\\
% ...
$^{101}$CPPM, Aix-Marseille Universit\'e, CNRS/IN2P3, Marseille; France.\\
$^{102}$Department of Physics, University of Massachusetts, Amherst MA; United States of America.\\
% ...
$^{128}$Homer L. Dodge Department of Physics and Astronomy, University of Oklahoma, Norman OK; United States of America.\\
% ...
$^{156}$Department of Physics and Astronomy, University of Sussex, Brighton; United Kingdom.\\
% ...
$^{161}$Raymond and Beverly Sackler School of Physics and Astronomy, Tel Aviv University, Tel Aviv; Israel.\\
% ...
$^{167}$Department of Physics, University of Toronto, Toronto ON; Canada.\\
% ...

\end{flushleft}


\end{document}